\documentclass[graybox]{svmult}

\usepackage{type1cm}        
\usepackage{makeidx}         
\usepackage{graphicx}        
\usepackage{subcaption}
\usepackage{multicol}        
\usepackage[bottom]{footmisc}
\usepackage{caption}

\usepackage[utf8]{inputenc}
\usepackage[T1]{fontenc}

\usepackage{newtxtext}       %
\usepackage[varvw]{newtxmath}       

\usepackage{comment}
\usepackage{cite}

\usepackage{siunitx}

\DeclareSIUnit{\angstrom}{\mbox{\normalfont\AA}}

\makeatletter
\DeclareRobustCommand{\AA}{%
  \leavevmode
  \vbox{\ialign{##\cr
    \hidewidth\char'27 \hidewidth\cr
    \noalign{\nointerlineskip\kern-1.4ex}
    A\cr
  }}%
}
\makeatother

\newcommand{\R}{\mathbb{R}}

\DeclareMathOperator*{\argmin}{argmin}

\makeindex             

\begin{document}

\title*{On the Interplay of Subset Selection and Informed Graph Neural Networks}
\author {Niklas Breustedt$^\star$, Paolo Climaco$^{\ddagger}$, Jochen Garcke$^{\dagger \ddagger}$, Jan Hamaekers$^{\dagger}$, Gitta Kutyniok$^{+}$$^{\#}$, Dirk A. Lorenz$^\star$, Rick Oerder$^{\dagger}$,  and Chirag Varun Shukla$^{\#}$.}
\authorrunning {Breustedt, Climaco, Garcke, Hamaekers, Kutyniok, Lorenz, Oerder, Shukla}



\institute{$^{\dagger}$Fraunhofer SCAI, Sankt Augustin, Germany
\and $^{\ddagger}$Institut für Numerische Simulation, Universität Bonn, Germany 
\and $^{+}$University of Tromsø, Norway
\and $^{\#}$Ludwig-Maximilians-Universität München, Germany 
\and $^{\star}$Institut für Analysis und Algebra, Technische Universität Braunschweig, Germany}
%
%
\maketitle

\abstract*{Each chapter should be preceded by an abstract (no more than 200 words) that summarizes the content. The abstract will appear \textit{online} at \url{www.SpringerLink.com} and be available with unrestricted access. This allows unregistered users to read the abstract as a teaser for the complete chapter.
Please use the 'starred' version of the \texttt{abstract} command for typesetting the text of the online abstracts (cf. source file of this chapter template \texttt{abstract}) and include them with the source files of your manuscript. Use the plain \texttt{abstract} command if the abstract is also to appear in the printed version of the book.}

\abstract{Machine learning techniques paired with the availability of massive datasets dramatically enhance our ability to explore the chemical compound space by providing fast and accurate predictions of molecular properties. However, learning on large datasets is strongly limited by the availability of computational resources and can be infeasible in some scenarios. Moreover, the instances in the datasets may not yet be labelled and generating the labels can be costly, as in the case of quantum chemistry computations. Thus, there is a need to select small training subsets from large pools of unlabeled data points and to develop reliable ML methods that can effectively learn from small training sets. This work focuses on predicting the molecules’ atomization energy in the QM9 dataset. We investigate the advantages of employing domain knowledge-based data sampling methods for an efficient training set selection combined with informed ML techniques. In particular, we show how maximizing molecular diversity in the training set selection process increases the robustness of linear and nonlinear regression techniques such as kernel methods and graph neural networks. We also check the reliability of the predictions made by the graph neural network with a model-agnostic explainer based on the rate-distortion explanation framework.}

\section{Introduction}
Modelling the relationship between molecules and their properties is of great interest in several research areas, such as computational drug design \cite{Ma2015}, material discovery \cite{Mueller2016} and battery development \cite{Barker2021}. The field of computational chemistry offers powerful \textit{ab initio} methods to compute physical and chemical properties of atomic systems. 

Unfortunately, these approaches are often limited by their high computational complexity, which restricts their practical applicability to only small sets of molecules.
Therefore, machine learning (ML) methods for molecular property prediction have recently gained increased attention in molecular and material science because of their computational efficiency and accuracy on par with established first principle methods \cite{Gabor2021, Bochkarev2022, Batzner2022}. However, to effectively employ ML in real-world problems, there is a need for labelled datasets that can effectively represent the chemical space of interest, i.e., sets of molecules for which the target properties have already been computed using ab initio methods. Thus,  on the one hand, accurately choosing which data points to label in the analyzed chemical space is crucial to avoid creating a dataset with redundant information and limiting the required amount of ab initio calculations. On the other hand, it is critical to develop data-efficient ML methods that perform accurate predictions.

Integrating domain knowledge of physical and chemical principles into the dataset selection process and the development of  ML techniques is a primary goal of the chemical and material science ML community \cite{Kulik_2022}. Physical and chemical principles, such as spatial invariances, symmetries, algebraic equations and chemical properties, can increase the robustness, reliability and effectiveness of ML methods while reducing the required training data \cite{TaxonomyML2021, brandstetter2022geometric}. 

This work focuses on predicting the atomization energies of molecules in the QM9 dataset \cite{ramakrishnan2014quantum, Ruddigkeit2012} and shows how to exploit domain knowledge to select training sets according to specific criteria and how different ML methods may benefit from training on sets selected through such criteria. Specifically, by using Mordred \cite{Mordred}, a publicly available library, we generate knowledge-based vector representations of molecules based on their SMILES representation \cite{SMILES} without requiring any ab initio computations. Further, based on such a molecular vector-based representation, we define a training set selection process and can observe that a diversity in the selected subset can increase the reliability of ML methods, indicated by the reduction of the maximum absolute error of the prediction. The maximum absolute error can be interpreted as a measure of robustness, and it is a helpful metric to evaluate ML methods in chemical and material science \cite{Zaverkin2022} since the average error alone gives an incomplete impression~\cite{D2CP00268J,Sutton2020}.
Furthermore, this work shows how diversity reduces the gap between the predictive robustness of linear regression-based approaches relying only on the molecular topological information, such as kernel ridge regression (KRR)\cite{Kung2014}, and non-linear approaches relying on molecular geometric representations obtained through ab initio computations, such as graph neural networks (GNN) \cite{jorgensen2018, kipf2017, gilmer2017}. We compare the effectiveness of a diversity-based selection with that of random sampling and of an alternative selection approach based on domain knowledge that focuses on representativeness, i.e., the distribution of chosen properties of the dataset should be present with the same amount in the selected training sets.

Finally, we note that our GNNs are inherently opaque (i.e. the logic flow to the decision-making process of the neural network is obscured). This inherently opaque nature of common deep neural network architectures has led to a rise in demand for trustworthy explanation techniques, which vary in their meaning and validity~\cite{Roscher.Bohn.Duarte.ea:2019}. Unlike other modalities in computer vision and natural language processing, the non-Euclidean nature of graph-structured data poses a significant challenge to trustworthy and interpretable explanation generation. To this end, there exist a variety of explanation techniques and explanation types\cite{graphlime,graphmask,graphsvx,subgraphx,gnnexplainer,xgnn, pgex,pope}, the most popular of which are subgraph explanation techniques. 

We probe the domain knowledge learned/retained by our GNNs for different sampling strategies through the application of a novel \textit{post-hoc} model-agnostic explanation technique, graph rate-distortion explainer (GRDE). GRDE builds on the existing rate-distortion explanation (RDE) framework \cite{kolek,macdonald2019} to generate \textit{instance-level} subgraph explanations on the input graphs, which highlight the substructures and features in the graph that are most relevant towards the GNNs' predictions.

In the following, we first give an overview on three ML models that are designed for the prediction of molecular properties but are based on different underlying working principles. In this way, we hope that our results yield insights for a variety of methods that are used in practical applications. Following that, we discuss two ways of sampling subsets from a larger dataset, one aiming to maximize the diversity of the selected samples and the other seeking to choose a collection of points representative of the set from which we sample. Afterwards, we test the introduced methods, namely the SchNet, KRR and the spatial 3-hop convolution network which is proposed in this work, by performing numerical experiments on the QM9 dataset while putting special emphasis on the effects of the sampling strategies. After a discussion of the numerical results and a comparison between the different ML models, we seek explanations of the model predictions by applying GRDE to one of the employed graph neural networks.

\section{Related Work}
 In recent years, there has been growing interest in incorporating domain-specific knowledge into the selection of training data and the development of learning algorithms, which is referred to as informed machine learning \cite{TaxonomyML2021}. Ideally, the training data selection process should be based purely on the data's features, as labels may be expensive to compute, and should be model-independent so that the selected training data is beneficial for multiple learning models rather than just one. This allows for greater flexibility in model selection and avoids the need for repeating the dataset selection process for each model. Considering these practical aspects, it is clear that a feature-based and model-independent selection process is desirable for efficient and effective machine learning. This section reviews some of the relevant work in this area. 
Coreset approaches \cite{Feldman2019} are among the most popular strategies for feature-based and model-independent selection of training datasets. Several of these approaches involve incorporating domain-knowledge into the selection of training data by selecting data points that are representative of the distribution of the target points for which we want to predict the new labels. The simplest and yet one of the most common coreset approaches is uniform sampling, which involves selecting a random subset of data points from the larger dataset. Uniform sampling is also considered a benchmark for every other selection approach. Unfortunately, uniform sampling does not exploit domain knowledge and can lead to biased results if the dataset is imbalanced or if certain data points are more important than others. To address this issue, importance sampling \cite{Braverman2016NewFF} is an approach that exploits domain knowledge to assign weights to each data point based on its importance or relevance to the problem at hand. The weights are then used for a nonuniform selection of the training set that privileges more important data points. Another class of methods are the grid-based approaches \cite{Agarwal2005}, which involve dividing the feature space into a grid and selecting one or more representative points from each grid's cell. This can be useful for problems with a high-dimensional feature space or when there is a need for a more structured selection of data points. Greedy constructions are coreset approaches that iteratively select the most informative data points based on a pre-defined criterion. For instance, well-known greedy selection methods are submodular function maximization algorithms \cite{krause2014submodular}. Greedy approaches can be effective for selecting a small subset of highly informative data points, but they may be computationally expensive for large datasets. Overall, the choice of coreset approach depends on the specific problem and dataset characteristics, as well as computational constraints. See \cite{Feldman2019} for a more detailed review of coreset approaches. Finally, the field of experimental design~\cite{Yu06} offers additional sampling strategies to perform a feature-based selection of the training set that can benefit specific regression model classes, e.g., linear models.

In this work, incorporating domain knowledge in the learning of algorithms refers to methods which are known as informed graph neural networks. While graph neural networks recently gained increasing attention by the works from Gori et al.~\cite{gori2005} and Scarselli et al.~\cite{scarselli2009}, the question of how to use domain knowledge to improve the performance of learning methods dates back to the last century (e.g. see~\cite{joerding1991} or~\cite{kramer1992}). More recently, physics informed neural networks, which address supervised learning tasks complying with the known laws from physics, are a hot topic in several applications, e.g. to find surface breaking cracks in a metal plate~\cite{shukla2020} or to solve inverse heat transfer problems~\cite{cai2021}. For graph neural networks, based on the message passing principle, i.e. the process of updating so called states or representations attached to each node of a graph using the node's neighbourhood, many different models were proposed (e.g. Graph Attention Networks~\cite{velivckovic2017}, ChebNet~\cite{defferrard2016}, Gated Graph Neural Networks~\cite{li2015}), the most popular being the graph convolutional model by Kipf and Welling~\cite{kipf2017} which is motivated by an approximation of spectral graph convolutions. Combining incorporating domain knowledge with graph neural networks leads to the very recent informed graph neural networks. In~\cite{hernandez2022} the authors combine theory from thermodynamics with graph neural networks to predict the behaviour of dynamical systems and in~\cite{kim2022} combine physical properties of molecules are combined with graph neural networks to predict the cetane number of possible alternative fuels. 
For more detailed overviews on GNNs or informed neural networks we refer to the book~\cite{liu2020} and a recent review~\cite{cuomo2022}. 

We further build upon the interpretability of graph neural networks in this work by introducing a method akin to perturbation techniques on image data to graph-structured data. The main goal of interpretability is to invoke transparency in the otherwise opaque prediction process of neural networks, and is further applicable in the detection of bias as well as to explain incorrect classifications in the predictive model. Previous work in interpretability for other modalities such as audio and images~\cite{kolek,macdonald2019} has shown great success in identifying a neural network's sensitivity to specific subsets of the input data. More specifically, among the variety of local and global interpretability techniques, perturbation~\cite{kolek,kolek2022explaining} and gradient-based~\cite{smilkov2017smoothgrad} techniques have been shown to accurately capture a predictive model's sensitivity to some concepts in the input data. These techniques generally seek to optimize a heatmap over the input data such that high-intensity zones are the most relevant to the model's prediction for the given data point. We further discuss this in detail with respect to graphs in section \ref{subsec: GRDE}. 

Inspired by the exhaustive work on interpretability for other modalities, several methods\cite{pgex,gnnexplainer, pope,graphmask} have also been proposed for graph-structured data, with perturbation techniques such as GNNExplainer~\cite{gnnexplainer} being the baseline for comparison. For a detailed overview of GNN interpretability, we refer to~\cite{yuan2022explainability}. These techniques, however, have been shown to suffer from unfaithfulness on large graphs since they optimise masks only for small graphs as well as manually threshold their relevance scores. See ~\cite{agarwal2023evaluating} for a detailed review on the current issues with graph interpretability. 

\section{Methods and Sampling Strategies}
\label{Methods and Sampling Strategies}
This section introduces the approaches we use for predicting the atomization energy, explaining the GNN output and sampling the training data. Subsection \ref{subsec:SchNet} introduces the benchmark regression model SchNet, a GNN that uses 3-dimensional positional information to predict chemical properties. Next, subsections \ref{subsec:GaussianKernelRegression} and \ref{subsec:ThreeHop} describe KRR and the spatial 3-hop convolution network, respectively. Both these approaches only exploit topological information encoded in the SMILES to perform the energy prediction task. Subsection \ref{subsec: GRDE} presents the rate-distortion explanation framework for graph data that we use to showcase the domain knowledge learned by the 3-hop convolution network. Finally, subsection \ref{sampling_strategies} introduces the approaches we use for the selection of training sets.

\subsection{SchNet}
\label{subsec:SchNet}

SchNet is a symmetry-informed neural network model, designed for the prediction of chemical properties by Schütt et al. \cite{Schuett2017}. In contrast to the methods presented in sections \ref{subsec:GaussianKernelRegression} and \ref{subsec:ThreeHop}, it is trained and evaluated on 3-dimensional structural information describing the atomic systems of interest. Usually, the positional information is obtained from computational methods such as density functional theory (DFT). 

More formally, for an atomic system with N atoms, SchNet can be used to predict scalar properties as a function $f$ of $3N$ atomic coordinates (nuclear positions) and on $N$ atomic numbers of the corresponding atoms: 

\begin{equation}
    f: \mathbb{R}^{3N} \times \mathbb{N}^N \xrightarrow{} \mathbb{R}.
\end{equation}

Internally, SchNet operates on a distance-based neighborhood graph, defined by a cutoff radius $r_\mathrm{cut}$, in which nodes correspond to the atoms in the atomic system. In this scenario, edges do not necessarily correspond to chemical bonds but merely indicate whether two atoms are closer than the chosen cutoff radius. Hence, the chosen cutoff radius has a direct influence on the graph shown to the model. Similar to other GNNs \cite{gilmer2017, kipf2017}, SchNet operates in a layer-wise fashion by iteratively updating feature representations. At the $l$-th layer each atom, indexed by $i \in \{1,2,...,N\}$,  is represented by a feature vector $\pmb{x}_i^l \in \mathbb{R}^F$ where $F$ is a hyperparameter. The main layer introduced by Schütt et al. is the continuous-filter convolutional layer: Denoting the atomic positions by $\pmb{r}_i \in \mathbb{R}^3$, this layer updates the atomic features as follows:

\begin{equation}
    \pmb{x}^{l+1}_i = \sum_{j \in \mathcal{N}(i)} \pmb{x}^l_j \circ W^l \left(\pmb{r}_i - \pmb{r}_j\right),
\end{equation}
where $W^l: \mathbb{R}^3 \xrightarrow[]{} \mathbb{R}^F$ is a trainable filter-generating function and $\circ$ denotes element-wise multiplication. In detail, $W^l$ is given as the composition $W^l = \tilde{W}^l \circ \varphi$ of a distance-based radial basis expansion 
\begin{equation}
    \varphi: \pmb{r}_i - \pmb{r}_j \mapsto \bigoplus_{k=1}^{N_\mathrm{radial}}  \exp \left( - \gamma \left( \lVert \pmb{r}_i - \pmb{r}_j \rVert_2 - \mu_k \right)^2 \right)
\end{equation}
and a trainable neural network $\tilde{W}^l$ where $\SI{0}{\angstrom} \leq \mu_k \leq \SI{30}{\angstrom}$ are equidistributed centers and $\gamma=\SI{10}{\angstrom}$. Here, $\bigoplus$ denotes the direct sum that concatenates the scalar outputs of the radial basis functions to a feature vector in $\mathbb{R}^{N_\mathrm{radial}}$ which is then passed into $\tilde{W}^l$.
Note that $\varphi$ is invariant with respect to actions of the orthogonal group $O(3)$ which assures that the predictions of SchNet are invariant with respect to translations, rotations and reflections of the input structure as well. Depending on the atomic species, initial embeddings $\pmb{x}^0_i$ are sampled from an $F$-dimensional standard normal distribution and optimized during the training process. In addition, non-linear layers such as dense feed-forward neural networks can be applied to the node features in order to increase the expressiveness of the model. 

By summing over the images of a trainable readout function $R: \mathbb{R}^F \xrightarrow[]{} \mathbb{R}$, the final node features in the last layer $L$ are transformed into a prediction of the target property $\hat{y}$:

\begin{equation}
    \hat{y} = \sum_{j=1}^N R \left( \pmb{x}_j^{L} \right)
\end{equation}

Involving only permutation-invariant operations such as the summation over adjacent atoms, the output is invariant with respect to mutual permutations of the atomic positions and atomic species. For more details on the model architecture see \cite{Schuett2017}.  

\subsection{Kernel Ridge Regression}
\label{subsec:GaussianKernelRegression}

In kernel ridge regression, a vector-based representation of the molecules is mapped into a high-dimensional space using a non-linear map that is implicitly determined by defining a kernel function, which provides a measure of similarity between the molecular representations. The structure-energy relationship is learned in the high-dimensional space. In this work, we use the so-called Gaussian kernel

\begin{equation}
   k(\pmb{x}_i, \pmb{x}_j) := e^{-\frac{\|\pmb{x}_i-\pmb{x}_j\|_2^2}{2 \nu ^2}},
\end{equation}
where $\| \cdot \|_2$ is the $L_2$-norm and $\nu \in \mathbb{R}$ a kernel
hyperparameter to be selected through an optimization process. 
The kernel ridge regression model is constructed using the selected training set $\{\pmb{x}_i,y(\pmb{x}_i)\}_{i=1}^p$, where $\{\pmb{x}_i\}_{i=1}^p$ are the Mordred \cite{Mordred} based vector representations of the molecules and $\{y(\pmb{x}_i)\}_{i=1}^p$ the associated atomization energies. Once the regression model has
been constructed, the predicted energies are given by the scalar values $\tilde{y}(\pmb{x})$ defined as follows
 \begin{equation}
   \label{kernel_prediction}
   \tilde{y}(\pmb{x}) := \sum_{i=1}^p\alpha_ik(\pmb{x},\pmb{x}_i),
 \end{equation}
where the vector $\pmb{\alpha}=[\alpha_1,
\alpha_2,\dots,\alpha_p]^T \in \mathbb{R}^p$ is the solution of the following
minimization problem
\begin{equation}
   \label{KRR_minimization}
   \pmb{\alpha} = \argmin_{\bar{\pmb{\alpha}}}\sum_{i=1}^p(\tilde{y}(\pmb{x}_i)-y(\pmb{x}_i))^2 + \lambda \bar{\pmb{\alpha}}^T\pmb{K} \bar{\pmb{\alpha}}.
\end{equation}
 Here, $\pmb{K}$ is the kernel matrix, i.e., $\pmb{K}_{i,j}= k(\pmb{x}_i,\pmb{x}_j)$, and the
 parameter $\lambda \in \mathbb{R}$ is the so-called regularization parameter that penalizes larger weights.
 The analytic solution to the minimization problem in (\ref{KRR_minimization})
 is given by
 \begin{equation}
   \pmb{\alpha}= (\pmb{K}+\lambda\pmb{I})^{-1}\tilde{\pmb{y}}
 \end{equation}
 where $\tilde{\pmb{y}}=[\tilde{y}(\pmb{x}_1),\tilde{y}(\pmb{x}_2),\dots,\tilde{y}(\pmb{x}_p)]^T$. Once the training process has been concluded and
 the regression parameters $\{\alpha_i\}_{i=1}^{p}$ have been learned, the energy predictions for
 molecules not included in the training set can be computed using Equation (\ref{kernel_prediction}).

\subsection{Spatial 3-Hop Convolution Network}
\label{subsec:ThreeHop}
In addition to the two previous approaches, we propose a third approach which builds on a newly developed spatial graph convolution structure. We call this approach spatial 3-hop convolution network. This approach exploits the graph structure, the node features and optionally edge features for regression or classification but does not need 3-dimensional structural information as is the case for SchNet. 

A commonly used graph convolutional network by Kipf \& Welling \cite{kipf2017}  is motivated by an approximation of a spectral convolution. Thereby, they consider spectral convolutions as 
 
 \begin{equation}
     \pmb{w} \star \pmb{x} = \pmb{U} \pmb{w} \pmb{U}^{\top}\pmb{x},
 \end{equation}
 where $\pmb{w} = diag(\theta)\in \R^{n\times n}$ is a filter, $\pmb{x} \in \R^{n}$ is a graph signal on a graph with $n$ nodes, $\star$ denotes the spectral graph convolution operator and $\pmb{U}$ is the matrix of eigenvectors from the eigendecomposition of the normalized graph Laplacian $\pmb{I}_{n} - \pmb{D}^{-\tfrac12}\pmb{A}\pmb{D}^{-\tfrac12}$. Moreover, $\pmb{A}$ is the adjacency matrix of the underlying graph, $\mathbf{D}$ is the corresponding degree matrix and $\pmb{I}_{n}$ is the $n \times n$ identity matrix. This convolution is approximated and generalized to matrix-valued graph signals which leads to the update of the graph convolutional network 
 
 \begin{equation}
     \pmb{H}^{(l+1)}= \sigma(\pmb{H}^{(l)}\pmb{W}_{0} + \pmb{D}^{-\tfrac12}\pmb{A}\pmb{D}^{-\tfrac12}\pmb{H}^{(l)}\pmb{W}_{1}),
 \end{equation}
 where $\pmb{H}^{(l)}$ is the matrix of hidden representations of the $l$-th layer, $\pmb{W}_{0}$ and $\pmb{W}_{1}$ are learnable parameters and $\sigma$ denotes the elementwise ReLU function.
  For the spatial 3-hop convolution layer we do not consider spectral graph convolutions but an intuitive spatial convolution using powers of the graphs adjacency matrix to calculate so called path matrices. Within these, for each node the number of paths of a certain length to every other node is stored. By defining a spatial convolution with path matrices and building a layer of the graph neural network using the convolution, we consider the number of paths of a given length from node $v$ to node $u$ as a measure for the impact of node $v$ on node $u$. Thus, nodes with more paths to the considered node will be taken into account more during the update.
  
  For a graph $G$ with $n$ nodes a path is defined as a sequence of nodes $(1,\ldots,k)$ with $k<n$ such that for any $i, j\in (1,\ldots,k)$ it is $i\neq j$, i.e. no node appears twice. 
 
 With that, we define a spatial $k$-hop graph convolution of a graph signal $\pmb{x} \in \R^{n}$ with a filter $\pmb{w} \in \R^{k}$ on an undirected graph $G$ with $n$ nodes as 
  \begin{align*}
     \pmb{w} \star_{k} \pmb{x} := \sum_{i = 0}^{k}{\pmb{w}_{i} \pmb{T}^{(i)}}x,
  \end{align*}
   where $\pmb{T}^{(i)}$ is a path matrix such that $\pmb{T}^{(i)}_{vu}$ is the number of paths of length $i$ from node $v$ to node $u$. 

 An approach to computing the needed path matrices is a recursion that starts with the adjacency matrix. Since the adjacency matrix equals the path matrix for paths of length one it is $\pmb{T}^{(1)} = \pmb{A}$. For every node $i$ and $u$ a neighbor of it, the number of paths of length two from node $i$ to node $j$ equals the number of paths of length one from $u$ to $j$ in which $i$ is not a part of. More generally, the number of paths of length $k$ from a node $i$ to a different node $j$ equals the sum of all paths from node $u$ to $j$ of length $k-1$ over all $u\in \mathcal{N}(i)$ in which $i$ does not appear. Using this, it can be shown that $\pmb{T}^{(2)} = \pmb{A}^{2} - \pmb{D}$ and $\pmb{T}^{(3)} = \pmb{A}^{3} - \pmb{\Sigma}\circ \pmb{A}$, where $\pmb{A}$ and $\pmb{D}$ are as above and $\pmb{\Sigma}$ is an $n\times n$ matrix with $\pmb{\Sigma}_{ij} = \pmb{D}_{ii} + \pmb{D}_{jj}$. This shows that the 3-hop spatial graph convolution is given by
  \begin{align*}
    \pmb{w}\star_{3}\pmb{x} = (\pmb{w}_{0}\pmb{I}_{n} + \pmb{w}_{1}\pmb{A} + \pmb{w}_{2}(\pmb{A}^{2}-\pmb{D}) + \pmb{w}_{3}(\pmb{A}^{3}-\pmb{\Sigma}\circ \pmb{A}))\pmb{x}.
  \end{align*}
  Note that the $\pmb{w}_k$'s can be seen as weights for the $k$-hop neighborhoods. 
  A generalization of the former discussion to a signal $\pmb{X}\in \R^{n \times d}$ with $c$ node features for each node (analogously to Kipf \& Welling \cite{kipf2017}) leads to 
  \begin{align*}
    \pmb{H} = \pmb{X}\pmb{W}_{0} + \pmb{A}\pmb{X}\pmb{W}_{1} + (\pmb{A}^{2}-\pmb{D})\pmb{X}\pmb{W}_{2} + (\pmb{A}^{3}-\pmb{\Sigma}\circ \pmb{A})\pmb{X}\pmb{W}_{3},
  \end{align*}
  which results in the spatial 3-hop convolution layer, the message passing layer of the spatial 3-hop convolution network,
    \begin{align*}
    \pmb{H}^{(l+1)}= \sigma(\pmb{H}^{(l)}\pmb{W}_{0} + \pmb{A}\pmb{H}^{(l)}\pmb{W}_{1} + (\pmb{A}^{2}-\pmb{D})\pmb{H}^{(l)}\pmb{W}_{2} + (\pmb{A}^{3}-\pmb{\Sigma}\circ \pmb{A})\pmb{H}^{(l)}\pmb{W}_{3}),
  \end{align*}
  where $\sigma$ is, again, the element-wise ReLU function. 
 
\subsection{Graph Rate-Distortion Explanations}
\label{subsec: GRDE}
We now present a formulation for the rate-distortion explanation framework~\cite{kolek,macdonald2019} for graph data.
Given a pre-trained GNN model, $\Phi : \mathbb{R}^{n\times c} \longrightarrow \mathbb{R}^m$ and a set of attributed graphs $G=\{G_1, G_2, ..., G_p\}$ such that 
$G_i=(V_i, E_i, X_i)$ for all $i \in [1, p]$, our task is to explain the model decision over the set $G$, or more locally, $\Phi(G_i)$. This leads us to the two general branches of explanation techniques: global and local explanations. Global explanation techniques focus on explaining the underlying function learned by the model, $\Phi$. This can be done in a multitude of ways, such as testing the model's sensitivity to a concept \cite{gcexplainer} or reconstructing graphs from the embedding space learned by the model to reveal important motifs \cite{xgnn}. In general, global explanation techniques, while useful, are hard to construct and are unable to detect finer details on local data points. On the other hand, local explanation techniques, which are the more popular alternative, focus on explaining $\Phi$ for local instances, i.e. $\Phi(G_i)$. Similar to global explanations, there exist a variety of approaches, such as perturbation-based methods \cite{pgex,gnnexplainer,graphmask}, surrogate methods \cite{graphlime}, gradient-based methods \cite{pope}, and additive methods\cite{graphsvx,subgraphx}, each with their benefits and limitations. These techniques aim to extract information from $G_i$ that is most relevant to the local prediction $\Phi(G_i)$. More concretely, given a graph $G_i=(A_i, X_i)$, local explanation techniques commonly attempt to extract a subgraph $\hat{G}_i=(\hat{A}_i,\hat{X}_i)\subseteq G_i$ that is  most relevant to the model for its prediction $\Phi(G_i)$. 
The rate-distortion framework for explaining graphs is a local, post-hoc, model-agnostic explanation technique that comes under the umbrella of perturbation-based graph explainers. Given the pre-trained model $\Phi$ and graph $G_i$, GRDE optimizes a binary deletion mask $S=(S_A, S_X)$ over $G_i$ to obtain a subgraph $\hat{G}_i$ such that $\Phi(\hat{G}_i)$ approximates $\Phi(G_i)$. Mask $S$ thus retains only the edges and features that are most relevant to the model's prediction on $G_i$.
Given $A_i\in\mathbb{R}^{n\times n}$ and $X_i\in\mathbb{R}^{n\times f}$, where $n$ is the number of nodes and $f$ is the number of node features, our goal is to optimize masks $S_A\in [0,1]^{n\times n}$ and $S_X\in [0,1]^{n\times f}$. Let $\mathcal{V}_S=(\mathcal{V}_{S_A}$, $\mathcal{V}_{S_X})$ be probability distributions that can either be chosen manually or learned from the graph dataset. Then the obfuscation on $G_i$, i.e. the subgraph $\hat{G}_i$, can be defined as
\begin{equation}
    \label{rde:obfuscation}
    \hat{G_i}=(\hat{A}_i,\hat{X}_i)=(A_i\odot S_A+(1-S_A)\odot \varv_{S_A}, X_i\odot S_X+(1-S_X)\odot \varv_{S_X}),
\end{equation}
where $\varv_{S_A}\in\mathcal{V}_{S_A}$, $\varv_{S_X}\in\mathcal{V}_{S_X}$, and $\odot$ denotes element-wise multiplication. Intuitively, this implies that the masks $S$ keep some of the elements in $G_i$ while the elements that are not selected by $S$ are replaced with values from the probability distribution $\mathcal{V}_{S}$ as 'noise'. In general, the choice of $\mathcal{V}_{S}$ should be such that the resulting subgraph $\hat{G}_i$ remains within the data manifold, provided that the data manifold is known. Depending on the information in $G_i$, we can use a variety of probability distributions for $(\mathcal{V}_{S_A}$, $\mathcal{V}_{S_X})$. For example, in the case of a binary adjacency matrix, $\mathcal{V}_{S_A}$ can be the Gumbel-Softmax distribution, whereas for real-valued adjacency matrices and node feature matrices, $\mathcal{V}_{S}$ can be Gaussian distributions. We can also learn the probability distributions $\mathcal{V}_{S}$ from the data manifold itself, as previous attempts have shown success with inpainting GANs\cite{kolek} for this strategy on other data modalities. 

Furthermore, we define the expected distortion on $G_i$ with respect to the masks $S$ and perturbation distributions $\mathcal{V}_S$ as
\begin{equation}
    \label{rde:distortion}
    \mathcal{D}(G_i, S,\mathcal{V}_S, \Phi) = \mathop{\mathbb{E}}_{\varv_{S_A}\in \mathcal{V}_{S_A}, \varv_{S_X}\in \mathcal{V}_{S_X}} \Big[d(\Phi(G_i),\Phi(\hat{G}_i))\Big],
\end{equation}
where $d:\mathbb{R}^m \times \mathbb{R}^m \longrightarrow \mathbb{R}_+$ is the measure of distortion between the two model outputs. Commonly, we can set $d$ as the $\mathcal{L}^2$ distance or the KL-divergence between the two model outputs. Thus, we can define the rate-distortion explanation on $G_i$ as the optimal subgraph $\hat{G}_i$ that solves the minimization problem 
\begin{equation}
    \label{rde: hard_dist}
    \mathop{min}_{S=(S_A,S_X)} \mathcal{D}(G_i, S,\mathcal{V}_S, \Phi) \text{\hspace{0.3cm} s.t \hspace{0.1cm}} ||S_A||_0\leq j, ||S_X||_0\leq k,
\end{equation}
where $j,k$ are the desired levels of sparsity for $S_A, S_X$ respectively. 

Note that solving equation (\ref{rde: hard_dist}) is $\mathcal{NP}$-hard\cite{macdonald2019}. Thus, we use an $l_1$ relaxation on equation (\ref{rde: hard_dist}) to get the relaxed optimization problem given by
\begin{equation}
    \label{rde:relax_dist}
    \mathop{min}_{S=(S_A,S_X)} \mathcal{D}(G_i, S,\mathcal{V}_S, \Phi) + \lambda_A||S_A||_1 + \lambda_X||S_X||_1,
\end{equation}
where $\lambda_A,\lambda_X>0$ are hyperparameters to control the sparsity level of the masks. We can further relax the binary masks $S$ by sampling them from the concrete distribution \cite{concrete} or Gumbel Softmax distribution \cite{gumbel_softmax}. This allows us to solve the optimization problem in equation (\ref{rde:relax_dist}) with differentiable techniques such as stochastic gradient descent. 

\subsection{Sampling Strategies}
\label{sampling_strategies}
We now introduce two approaches for sampling a set of points from a large dataset. The first method focuses on maximising the diversity of the
selected set, while the second aims to select a set that is
representative of the whole dataset.
\subsection*{Diversity}
In short, diverse subsets are iteratively selected from $\Omega \subset \mathbb{R}^d$ using the \textit{farthest point sampling} (FPS) algorithm~\cite{FPS}, where the resulting subset is a sub-optimal minimizer of the \textit{fill distance}. We denote this approach by FPS.

To maximize diversity of the selection we consider the concept of fill distance. Given a dataset $ \Omega \subset \mathbb{R}^d$
consisting of a finite amount of unique points, and $X = \{\pmb{x}_1,\pmb{x}_2,\dots, \pmb{x}_p \} \subset \Omega$ a subset of 
cardinality $p= |X| \in \mathbb{N}$ we define the fill distance of $X$ in $\Omega$ as
\begin{equation}
   h_{X, \Omega} :=\max_{\pmb{x} \in \Omega} \min_{\pmb{x}_j \in X}\| \pmb{x}-\pmb{x}_j\|_2.
\end{equation}
Put differently, we have that any point $\pmb{x} \in \Omega$ has a point $\pmb{x}_j \in X $ not farther away than $h_{X, \Omega}$.
Notice that, if $X, \bar{X} \subset \Omega$ with $p =|X| = |\bar{X}|$ and $h_{X, \Omega} < h_{\bar{X}, \Omega}$ then
$X$ consists of data points that are more widely distributed in $\Omega$, thus more diverse, than those in $\bar{X}$.

Fixing the number of points $p\in \mathbb{N}$ we want to select from $\Omega$, we aim to find $X \subset \Omega$ such that
\begin{equation}
   \label{Minimizing_fill_distance}
   X = \argmin_{\bar{X}\subset \Omega, |\bar{X}| = p} h_{\bar{X}, \Omega}.
\end{equation}
The naive approach to solve the minimization problem in
(\ref{Minimizing_fill_distance}) would first require computing the fill distance
for all possible sets $X \subset \Omega$ with $|X| =p$ and then choosing one of
those sets where the minimum of the fill distance is attained. Unfortunately,
such an approach is very time consuming and computationally intractable. Therefore, as an alternative approach we use the FPS algorithm \cite{FPS}. FPS is a greedy selection method, which means that the points are progressively selected starting from an initial a-priori chosen point, i.e., given a set of selected points
$X^s= \{\pmb{x}_1,\pmb{x}_2,\dots, \pmb{x}_s\} \subset \Omega$ with cardinality $ |X^s|=s <p$, the next chosen point is 
\begin{equation}
   \pmb{x}_{s+1}= \arg\max_{\pmb{x} \in \Omega}\min_{\pmb{x}_j \in X^s }\|\pmb{x}-\pmb{x}_j\|.
\end{equation}
$ \pmb{x}_{s+1}$ is the point which is farthest away from the points in $X^s$ and it is the point where the fill distance $ h_{X^s, \Omega}$ is attained. In other words, the next selected sample is the center of the largest empty ball in the dataset.

\subsection*{Representativeness}

We say that data points are representatively selected  for the entire dataset, when the distribution of properties in the selected subset are as close as possible to the corresponding distribution in the whole dataset. To this aim, we divide $\Omega$ into clusters and select data points from them
so that the distribution of the clusters in the subset resembles that of the whole dataset. For example, if we divide $\Omega$ into two clusters, each containing 50\% of the data points, we aim to select a subset consisting of data points which are also equidistributed in the two clusters. The clustering can be performed by clustering algorithms or be based on properties and criteria stemming from domain knowledge, i.e., in the sense of \cite{TaxonomyML2021} the training data is selected based on scientific knowledge. Furthermore, data points within each cluster are selected using the farthest point sampling, which ensures that in the various clusters a set of diverse data points is chosen. We call this approach cluster-based farthest point sampling (C-FPS).


\section{Numerical Experiments}

\subsection{QM9 Dataset} 
In this work, we analyze the publicly available QM9 dataset \cite{ramakrishnan2014quantum, Ruddigkeit2012} containing a diverse set of organic molecules. Precisely, the QM9  consists of 133 885  organic molecules in equilibrium with up to 9 heavy atoms of four different types: C, O, N and F.
The dataset provides the SMILES \cite{SMILES} representation of the relaxed molecules, their geometric configurations and 19 physical and chemical properties. 
To guarantee a consistent dataset, we remove all  3054 molecules that failed the consistency test proposed by \cite{ramakrishnan2014quantum}. Moreover, we remove the 612 compounds for which the RDKit package \cite{RDKIT} can not interpret the SMILES. After this preprocessing procedure, we obtain at a smaller version of the QM9 dataset consisting of 130219 molecules.

\subsubsection{Knowledge Based Molecular Representation}

The domain knowledge based molecular representation we employ is based on Mordred \cite{Mordred}, a publicly available library that exploits the molecules' topological information encoded in the SMILES strings to provide 1826 physical and chemical features. Such molecular features are defined as the ``final result of a logical and mathematical procedure, which transforms chemical information encoded within a symbolic representation of a molecule into a useful number or the result of some standardized experiment''\cite{todeschini2009molecular} and encode scientific knowledge reflecting algebraic equations, logic rules, or invariances~\cite{TaxonomyML2021}.  Using the Mordred library, we represent each molecule in the QM9 dataset with a high-dimensional vector where each vector's entry is associated with a distinct feature. 

To work with a more compact representation, after generating the Mordred vectors, we use the CUR \cite{Mahoney2009} approach to select a subset of relevant features. The CUR algorithm takes as input the Mordred vector representation of each of the molecules in the analyzed dataset and ranks the significance of the features by associating them with an importance score. We select the first 59 top-ranked features and normalize their values in the range (0,1) using the "MinMaxScaler" function provided by the scikit-learn python library \cite{scikit}.
Moreover, to ensure the uniqueness of the representation, we consider an additional set of features representing the atom type distribution within each molecule. Specifically, for each data point, we add five features, each expressing the amount of atoms of a particular type within the molecule, in percentage. The possible atom types are H, C, O, N and F. In conclusion, the Mordred based representation we employ to sample the QM9 dataset consists of 64-dimensional vectors.

\subsubsection{Diverse and Representative Sets of Molecules}

The knowledge related to the molecules in the QM9 enables us to employ the data sampling strategies introduced in subsection \ref{sampling_strategies} to create diverse and representative sets.

Diverse sets are constructed using the FPS algorithm on the Mordred-based vector representations of the molecules in the QM9. The Mordred vectors allow the representation molecules as points in $\mathbb{R}^d$, $d \in \mathbb{N}$, and the definition of a distance between the molecules, the Euclidean distance. Thus, we represent the QM9 as a finite set $\Omega \subset \mathbb{R}^d$ and use the FPS to sample from $\Omega$ a sub-optimal minimizer of the fill distance.

Representative sets are constructed using the procedure introduced in subsection \ref{sampling_strategies} consisting of segmenting the QM9 in clusters and then sampling from each cluster so that the distribution of the chosen molecules resembles that of the whole dataset. The segmentation procedure is based on the molecules' topological information and considers their size, atom types and bond types, which following~\cite{TaxonomyML2021}, reflects scientific knowledge in the selection of training data so that selected molecular properties are invariant per cluster. Specifically, we define the clusters through a process consisting of three main steps. In the first step, we split the molecules according to their sizes. As a result of this first step, we divided the QM9 dataset into 26 sets. After that, we separate each cluster obtained in the first step into subclusters defined by the different heavy atom types within the molecules. Overall, molecules in the QM9 consist of 4 heavy atom types. Thus, each molecule could consist of 15 different combinations of such atom types, e.g., a molecule can contain up to four distinct heavy atoms, and for each amount of distinct heavy atom types, various combinations are possible. After this second step, each of the initial 26 clusters is divided into 15 subclusters. The third and final step is further splitting the data points in each subcluster into different sets according to the various bond types present in each molecule. We consider four different bond types: single, double, triple and aromatic bonds. Thus, each of the subclusters is further divided into 15 distinct sets. As a result of this clustering procedure, we divide the QM9 in 5850 different clusters that account for molecular size, atom types and bond types. Molecules within the clusters are selected using the farthest point sampling, which ensures that in the various clusters, a set of diverse molecules is chosen.

\subsubsection{Sampling the QM9 Dataset}
For the experiments, we select training sets of different sizes and according to different strategies from the entire preprocessed QM9 dataset. After that, we test each trained model's predictive accuracy on all the molecules that have not been selected to train it.
We construct training sets consisting of 100, 250, 500, 1000 and 5000 samples. Such sets are created following three different selection criteria: random sampling (RDM), as a benchmark, and the two selection strategies introduced in section \ref{sampling_strategies}, namely, diversity sampling (FPS) and representative sampling (C-FPS).
For each sampling strategy and training set size, we run the training set selection process independently five times. For RDM, at each run the points are independently and uniformly selected, while in the case of FPS and C-FPS the initial point to initialize the FPS algorithm is independently selected at random at each run. Thus, for each selection strategy and training set size, each of the analyzed models is trained and tested five times, independently. The test results that follow are averaged over the five runs.

\begin{figure}[t]
\centering
\includegraphics[width=0.7\textwidth]{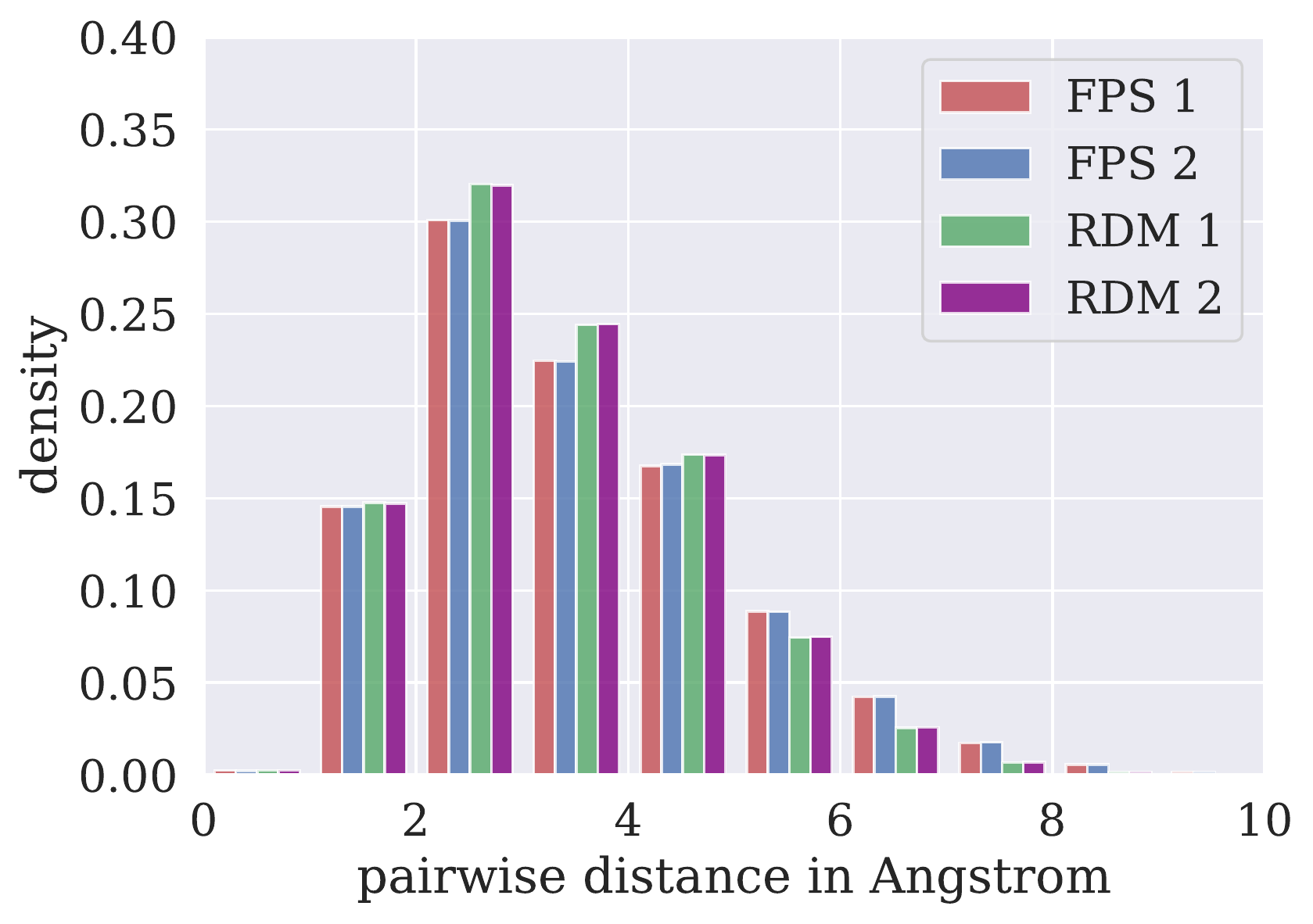}
\caption{The distributions of pairwise interatomic distances within a molecule for 5000 molecules sampled with either FPS or randomly (two different splits each) differ.}
\label{fig:pairwise_distances}
\end{figure}

We want to point out that sampling the training data non-randomly will lead to a shift between the training and test distribution, as showcased in Fig. \ref{fig:pairwise_distances}, where we compare FPS with a random selection. It is not obvious how such a bias effects the different models. Note that for Fig. \ref{fig:pairwise_distances}  we performed the selection twice, with different initialization for FPS and different seeds for the random selection, respectively.  We find that changing the initialization for FPS does not lead to a significant change in the distribution, for different seeds in the random selection we make the same observation.

\subsubsection{Measuring the Error}

We evaluate the performances of the employed machine learning methods using three different metrics. Specifically, we consider the mean absolute error (MAE), the root mean squared error (RMSE) and the worst-case error.
The mean absolute error (MAE) computes the arithmetic average of the absolute errors between the predicted values $\{\tilde{y}_i\}_{i=1}^N$ and the ground truths $\{y_i\}_{i=1}^N$, that is,
\begin{equation}
    \text{MAE} := \sum_{i=1}^N |y_i - \tilde{y}_i|,
\end{equation}
where $N \in \mathbb{N}$ is the number of data points in the test set used to evaluate the models.
The root mean squared error (RMSE) computes the root of the mean squared error, which is the arithmetic average of the squared errors. It is a measure of how spread out the errors are and it is represented by the formula 
\begin{equation}
    \text{RMSE} := \sqrt{\sum_{i=1}^N (y_i - \tilde{y}_i)^2}.
\end{equation}
The worst-case error calculates the maximum absolute error between the predicted values and the ground truths. It is an indicator of the robustness of a model's predictions, and it is defined as
\begin{equation}
\text{worst-case error} := \max_{1\leq i \leq N} |y_i - \tilde{y}_i|.
\end{equation}

\subsection{SchNet}

\begin{figure}[t]
\centering
\subfloat[MAE]{\label{fig:SchNet_results_a}\includegraphics[width=.49\linewidth]{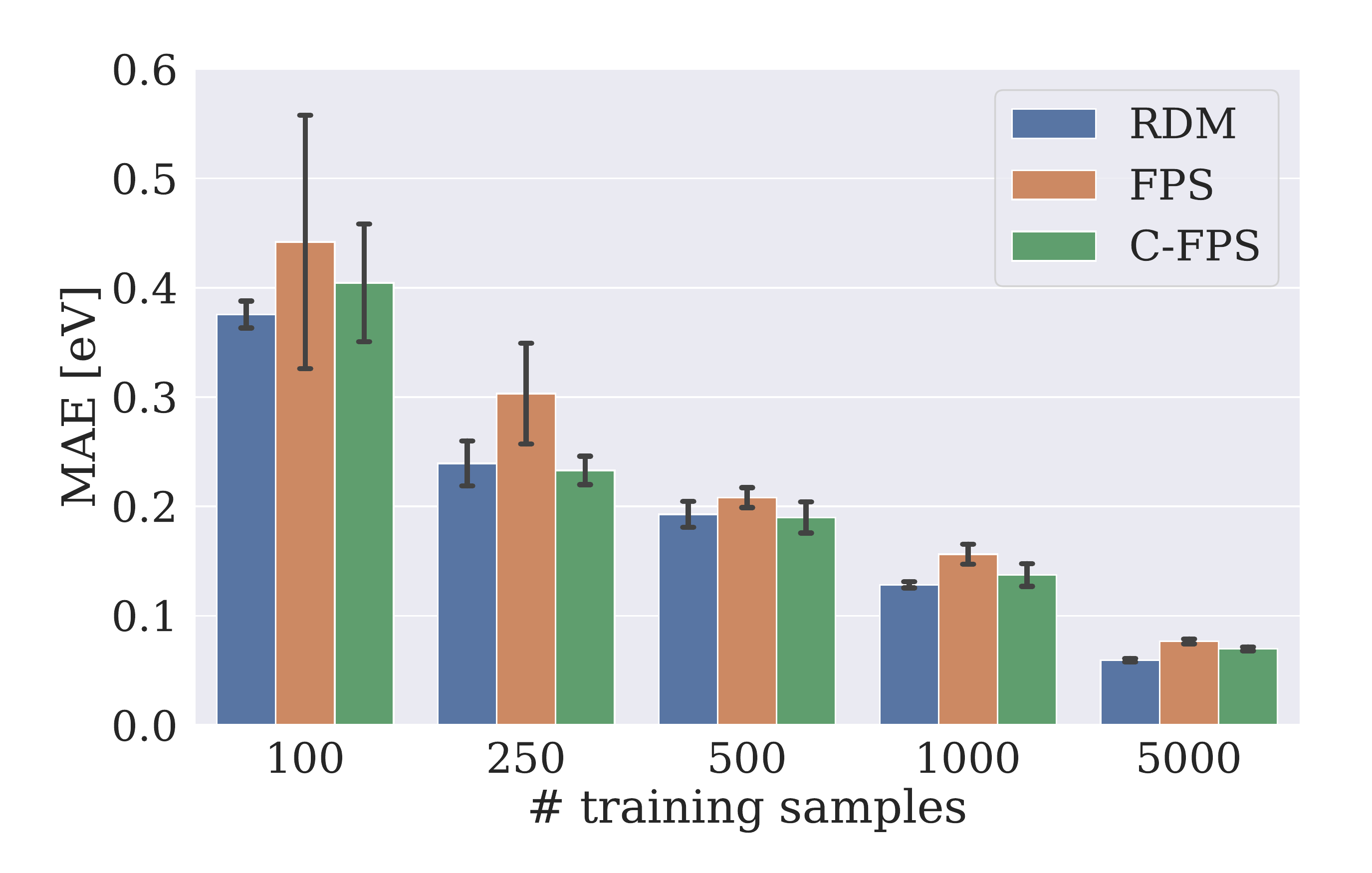}}\hfill
\subfloat[RMSE]{\label{fig:SchNet_results_b}\includegraphics[width=.49\linewidth]{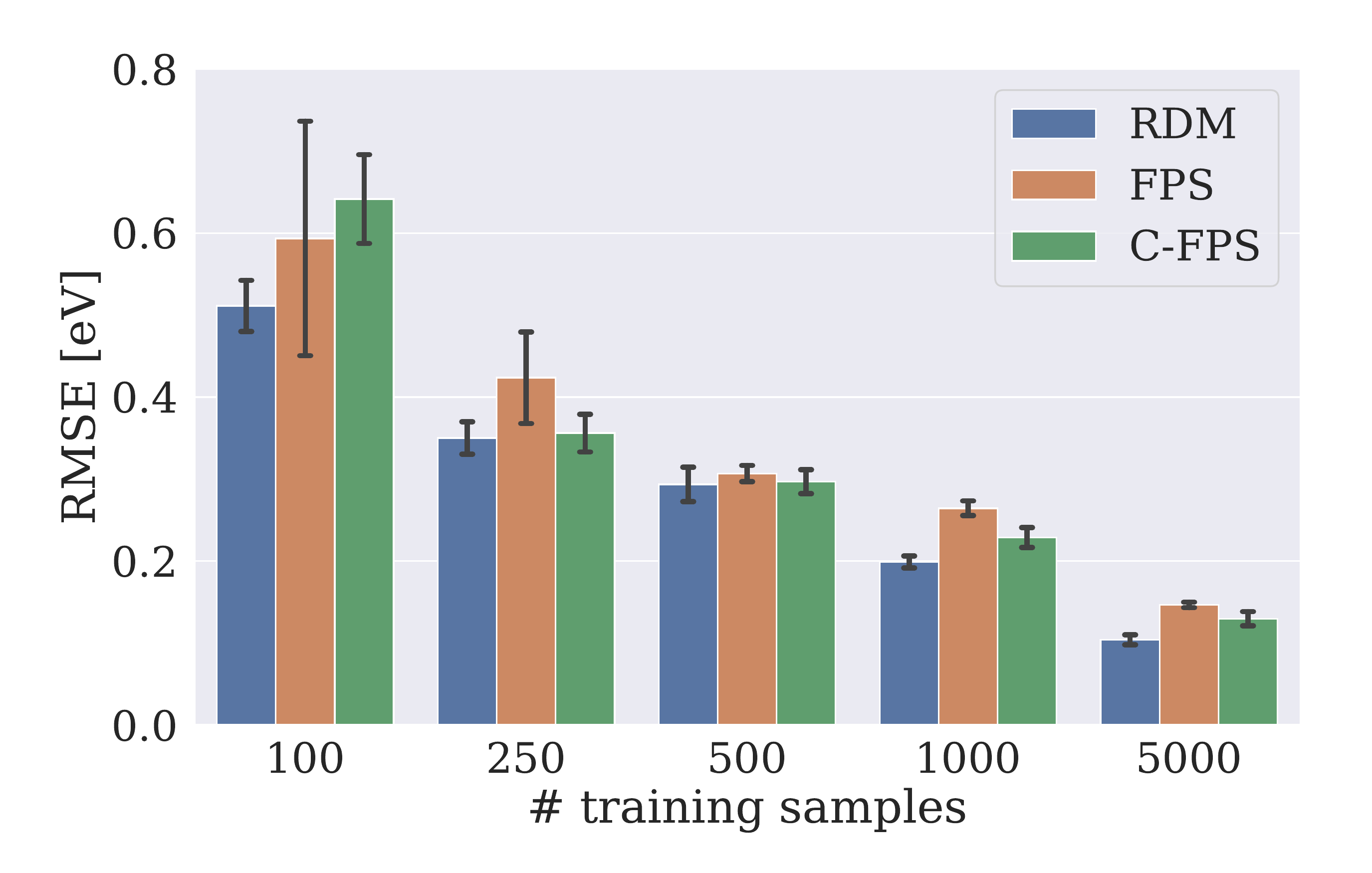}}\par 
\subfloat[Worst-case error]{\label{fig:SchNet_results_c}\includegraphics[width=.49\linewidth]{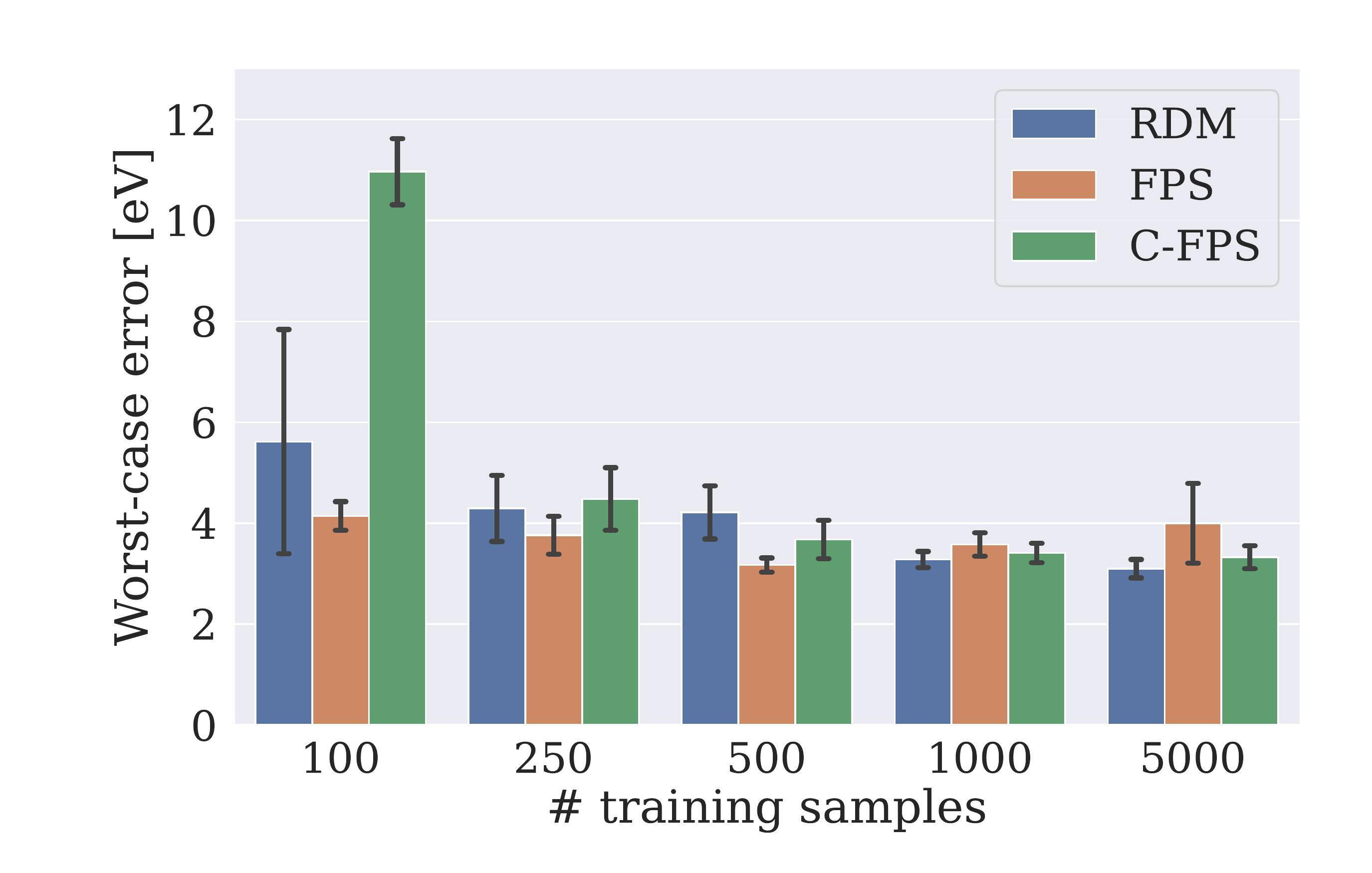}}
\caption{Results for SchNet.}
\label{fig:SchNet_results}
\end{figure}

In order to get experimental insights on FPS also for a different class of informed predictive models, we train the publicly available implementation of SchNet from Pytorch Geometric \cite{Fey2019} on the defined subsets. In this work, we choose a cutoff radius of $\SI{4}{\angstrom}$ while keeping the other hyperparameters to be the default ones suggested by the Pytorch Geometric implementation (version 2.0.4). Besides the test set that is used for final evaluation, we use random 20 \% from the training set for evaluation during training and refer to it as validation set. We minimize the $L2$-loss function with respect to the model parameters with the Adam optimizer using mini-batches of 32 molecules per iteration and a learning rate of $7 \cdot 10^{-4}$. The learning rate is decayed by a factor of 0.8 if the validation error has not improved for 50 epochs. After each epoch a checkpoint is saved if the model has achieved a smaller validation loss than the current best model. The training process is stopped after the model has not improved for 200 epochs (early stopping). The best model is then used for assessing the model performance on the test set. 

The first thing we observe is that SchNet does not seem to profit from FPS-based sampling strategies when examining the MAE and RMSE (Fig. \ref{fig:SchNet_results_a} and \ref{fig:SchNet_results_b}) alone. Random sampling consistently leads to approximately equal or smaller measurements for the MAE and RMSE for all investigated training set sizes. However, for 100, 250 and 500 training samples, the worst case error is reduced  by at least \SI{0.5}{\eV} when employing FPS for the training set selection (Fig. \ref{fig:SchNet_results_b}). Considering the comparatively small error bars, we expect FPS to be a reliable technique to reduce the worst case error for small (i.e. $\leq 500$ data points) training sets of QM9. For larger training sets however, this effect vanishes and FPS leads to worse results in the sense of larger worst-case errors. This is possibly due to the fact that FPS is based on Mordred features which yield a rather global description of a molecule. In this sense, FPS selects samples that are maximally far away with respect to those global features. On the contrary, GNNs strongly exploit local information and we believe this discrepancy to be a possible explanation for the merely small effect induced by FPS. However, in absolute numbers, SchNet yields the lowest error metrics of all tested methods. This meets our expectations since it is the only method incorporating geometric information. In fact, the nuclear positions were obtained through DFT calculations and hence the coordinates already encode highly relevant information for predicting the atomization energy. One could argue that SchNet's input is already part of the solution to the problem and view the use of features derived by ab initio methods as some form of information leakage \cite{gasteiger_gemnet_2021}, thus making the learning problem easier.

\subsection{Kernel Ridge Regression} 
The kernel and regression hyperparameters were optimized in a  grid search for each of the randomly selected training sets of 1000 points. Specifically, we varied the kernel parameter '$\nu$' in the set $\{10^{-1},10^{0}, 10, 10^{2},\dots, 10^{7}\}$ and the regularization parameter '$\lambda$' in the set $\{10^{-12},10^{-10}, 10^{-8}, 10^{-6},\dots, 10^{0}\}$. We selected $\nu= 10^5$ and $\lambda= 10^{-12}$, which is the parameter combination that provides the best performance in terms of the MAE on a randomly chosen test set consisting of 10000 points  not considered during training.

\begin{figure}[t]
\centering
\subfloat[MAE]{\label{fig:GaussianKernelRegression_results_a}\includegraphics[width=.49\linewidth]{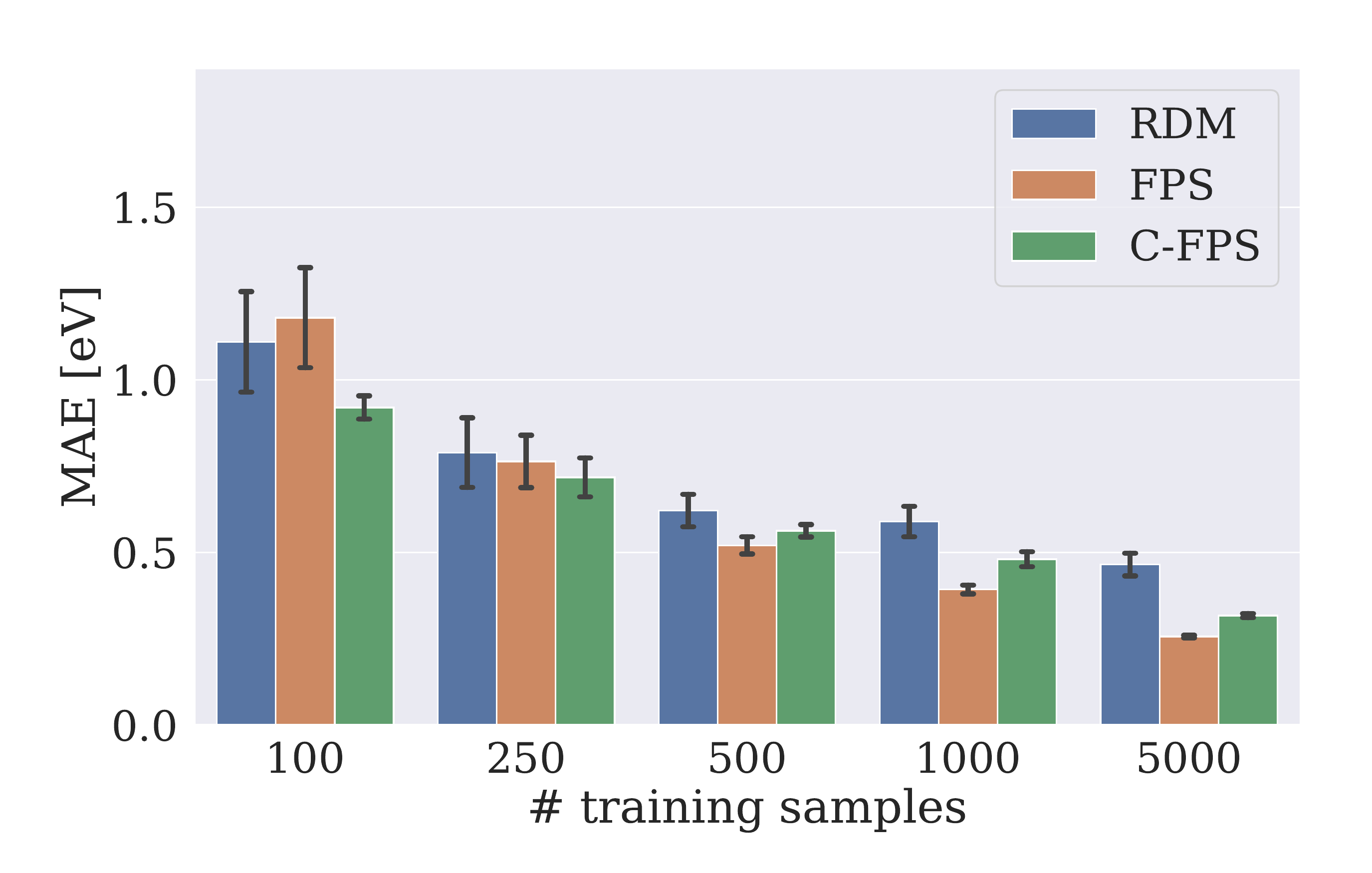}}\hfill
\subfloat[RMSE]{\label{fig:GaussianKernelRegression_results_b}\includegraphics[width=.49\linewidth]{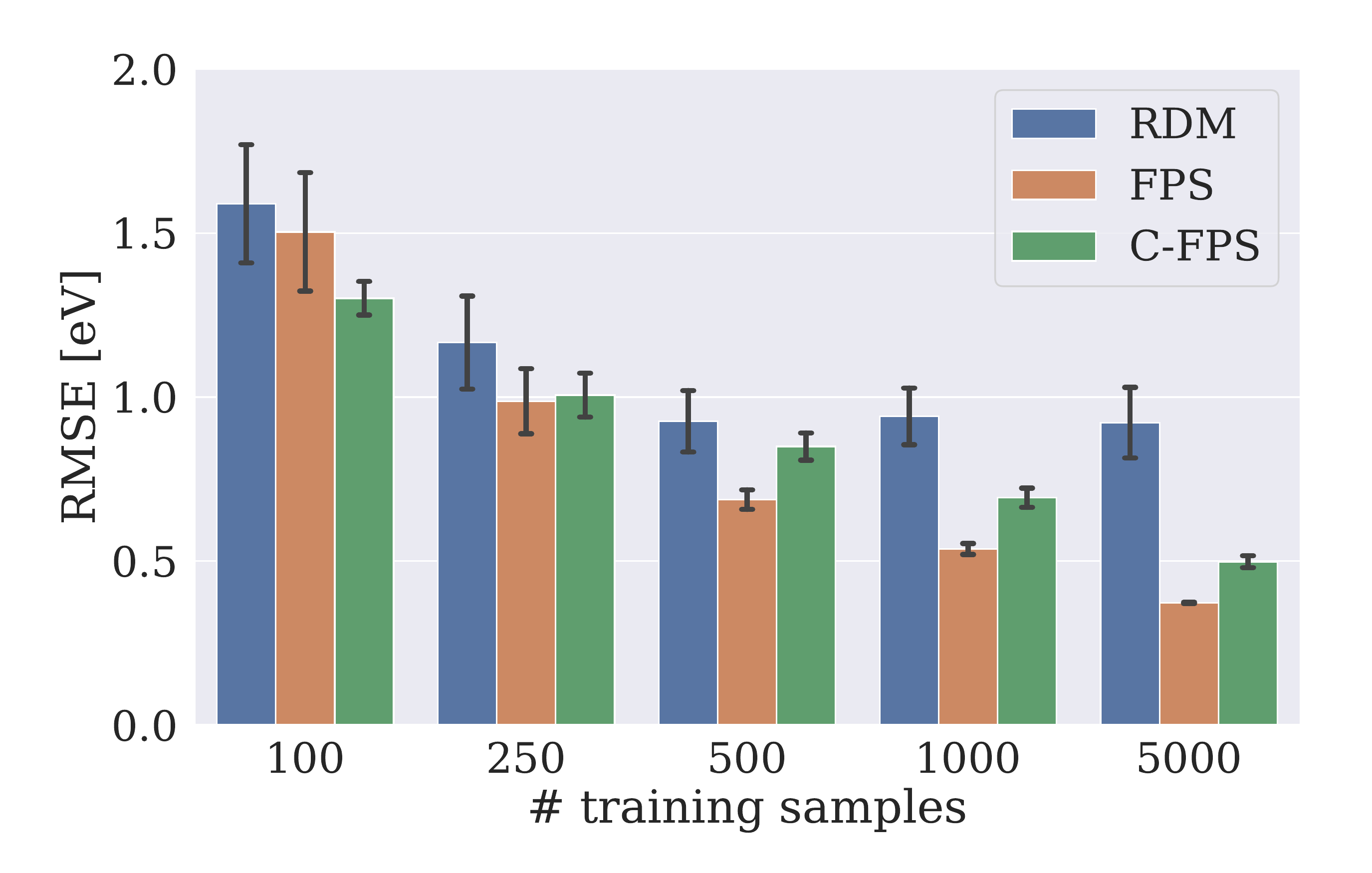}}\par 
\subfloat[Worst-case error]{\label{fig:GaussianKernelRegression_results_c}\includegraphics[width=.49\linewidth]{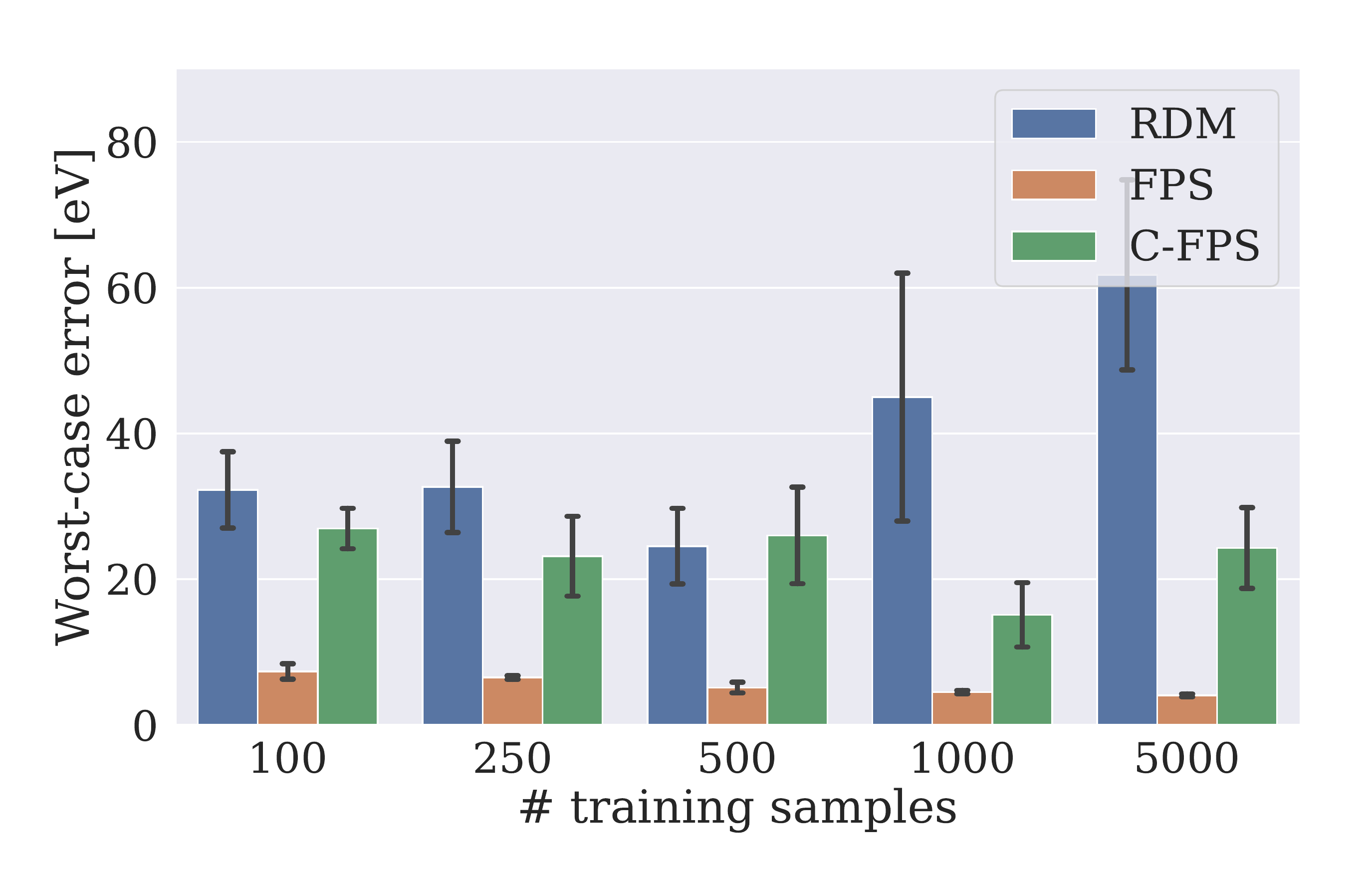}}
\caption{Results for Gaussian Kernel Regression.}
\label{fig:GaussianKernelRegression_results}
\end{figure}

Of all predictive models that we investigated, Gaussian kernel regression appears to benefit the most from FPS-based sampling strategies in comparison to random sampling. In particular, FPS and C-FPS improve the obtained RMSE on the test set for all training set sizes as seen in Fig. \ref{fig:GaussianKernelRegression_results_b}. However, it is noteworthy that random sampling leads to an increasing RMSE when going from 500 to 1000 or even 5000 training samples. For a possible explanation, we consider the MAE (Fig. \ref{fig:GaussianKernelRegression_results_a}) and the worst case error (Fig. \ref{fig:GaussianKernelRegression_results_c}). Even though we observe a decreasing MAE with an increasing training set size the worst case error becomes larger with more training samples as well, leading to a stagnating RMSE as it gives a higher weight to outliers than the MAE. FPS appears to alleviate this problem as becomes apparent when considering the comparatively small worst case errors. 

\begin{figure}[htb]
\centering
\includegraphics[width=0.7\textwidth]{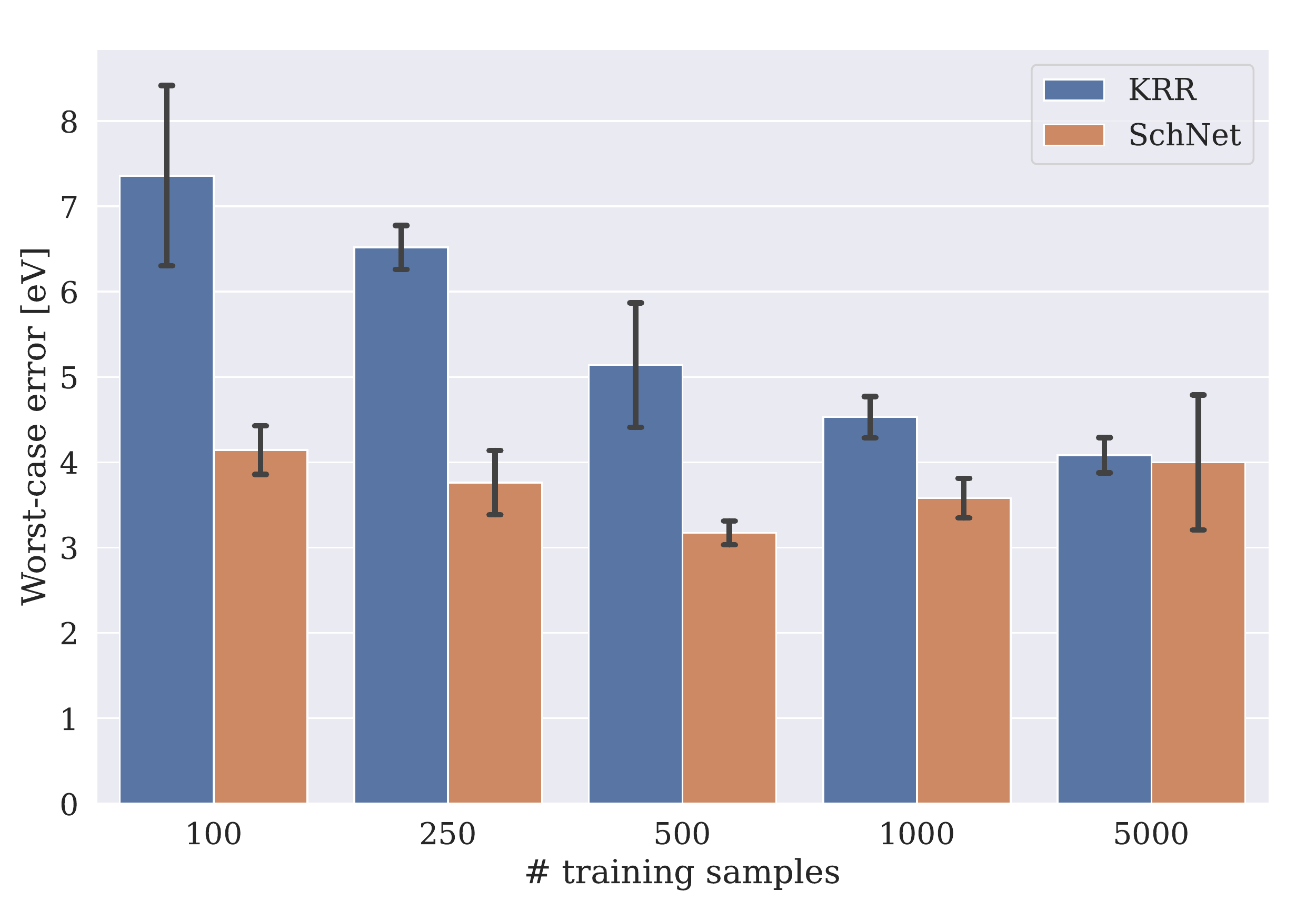}
\caption{The worst-case error of KRR  can be reduced by FPS such that the order of magnitude is comparable to SchNet.}
\label{fig:FPS_comparison}
\end{figure}

At this point, we can compare the worst case errors of the SMILES-based Kernel ridge regression (KRR) with the worst case error of SchNet. From Fig. \ref{fig:GaussianKernelRegression_results_c} it is apparent that FPS significantly reduces the worst-case error of KRR by one order of magnitude compared to random sampling. In order to contextualize this effect better we consider Fig. \ref{fig:FPS_comparison} that shows the worst-case errors of SchNet and KRR side by side for different numbers of training samples. We find KRR to approach the values of SchNet with an increasing number of training samples. In particular, we observe the errors to have the same order of magnitude. This is noteworthy as the KRR only exploits topological information while SchNet requires the atom coordinates obtained from DFT as input. 

\subsection{Spatial 3-Hop Convolution Network}

\begin{figure}[htb]
\centering
\subfloat[MAE]{\label{fig:Spatial_3_hop_convolution_network_results_a}\includegraphics[width=.49\linewidth]{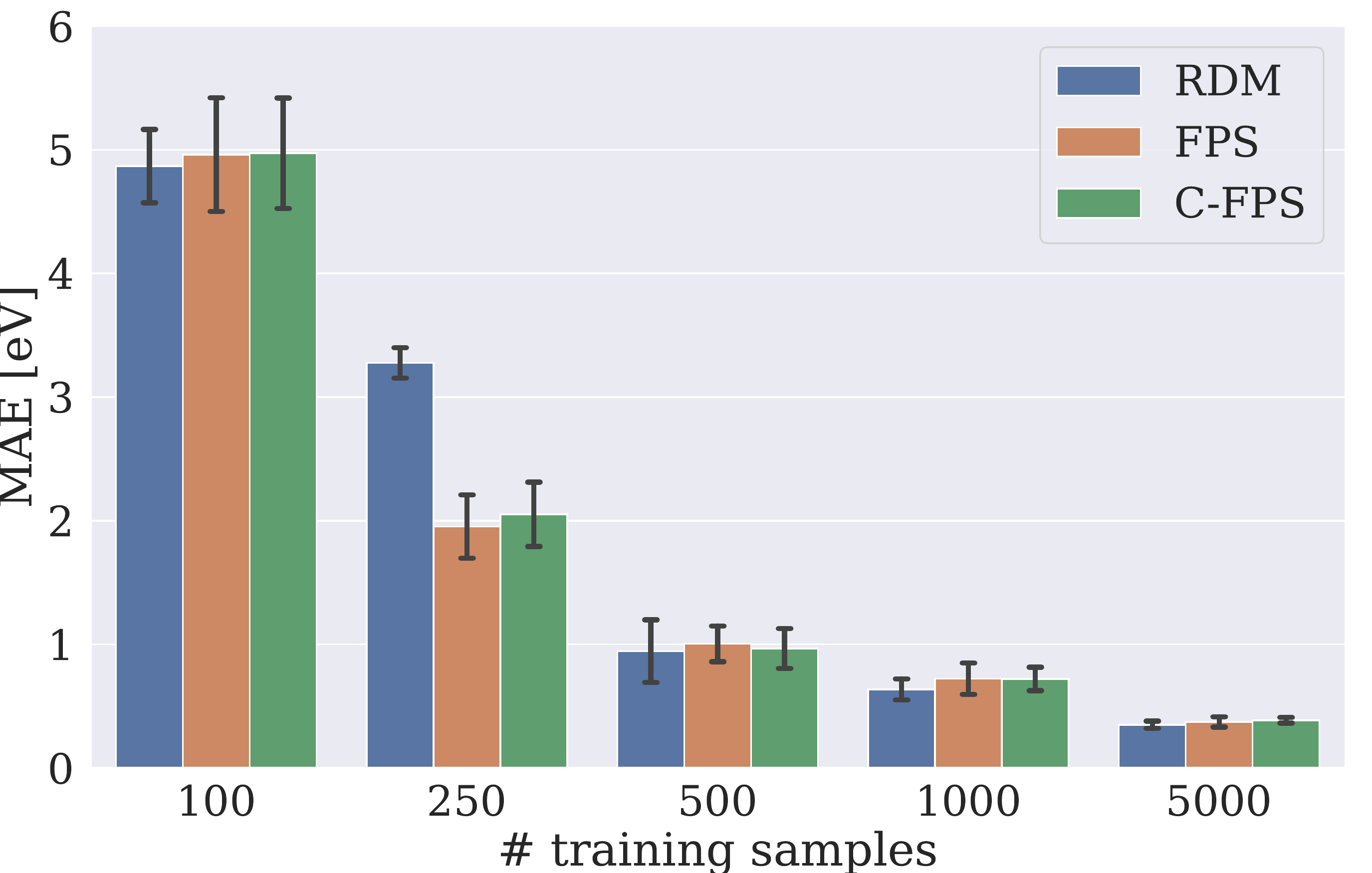}}\hfill
\subfloat[RMSE]{\label{fig:Spatial_3_hop_convolution_network_results_b}\includegraphics[width=.49\linewidth]{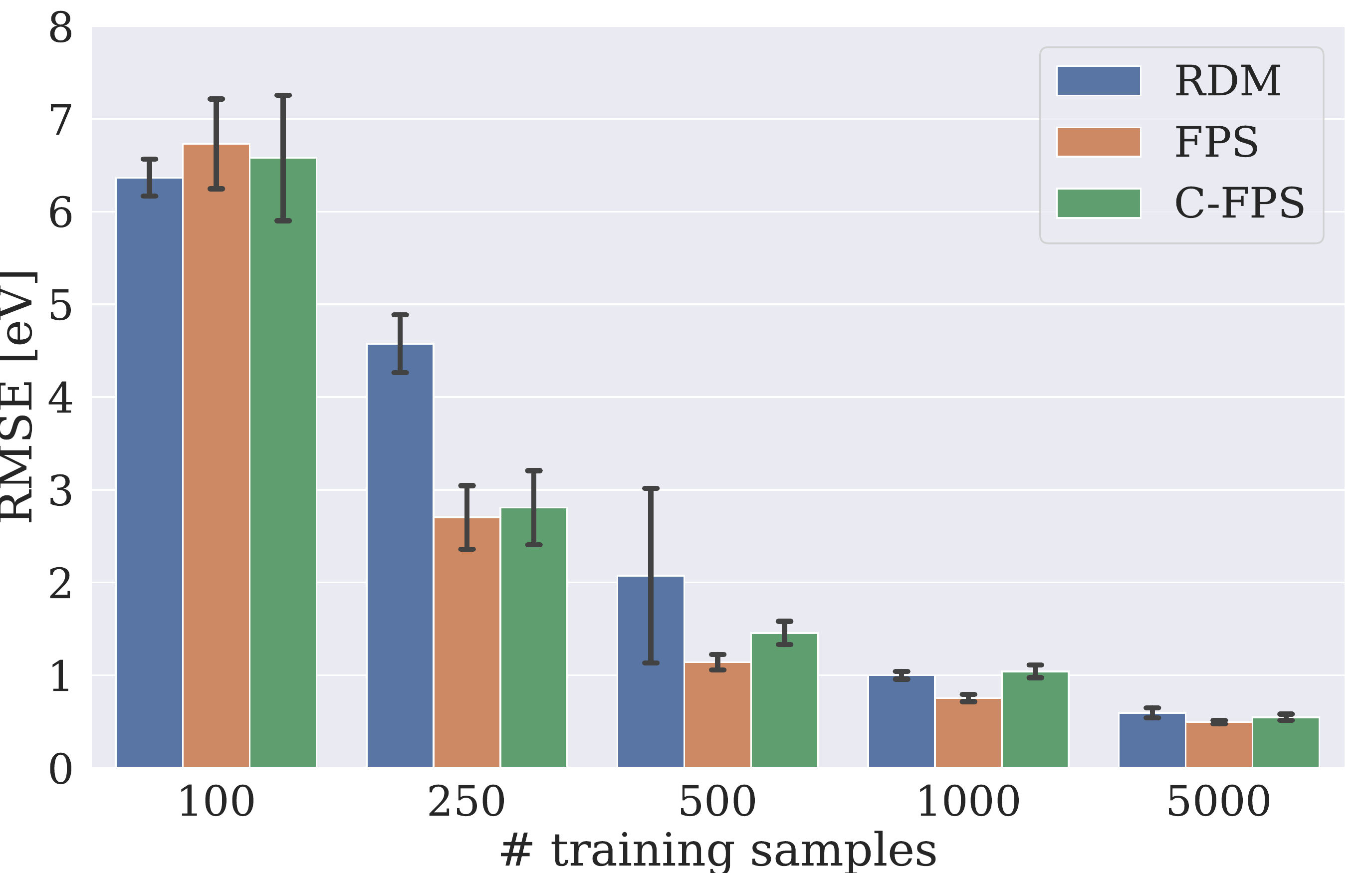}}\par 
\subfloat[Worst-case error]{\label{fig:Spatial_3_hop_convolution_network_results_c}\includegraphics[width=.49\linewidth]{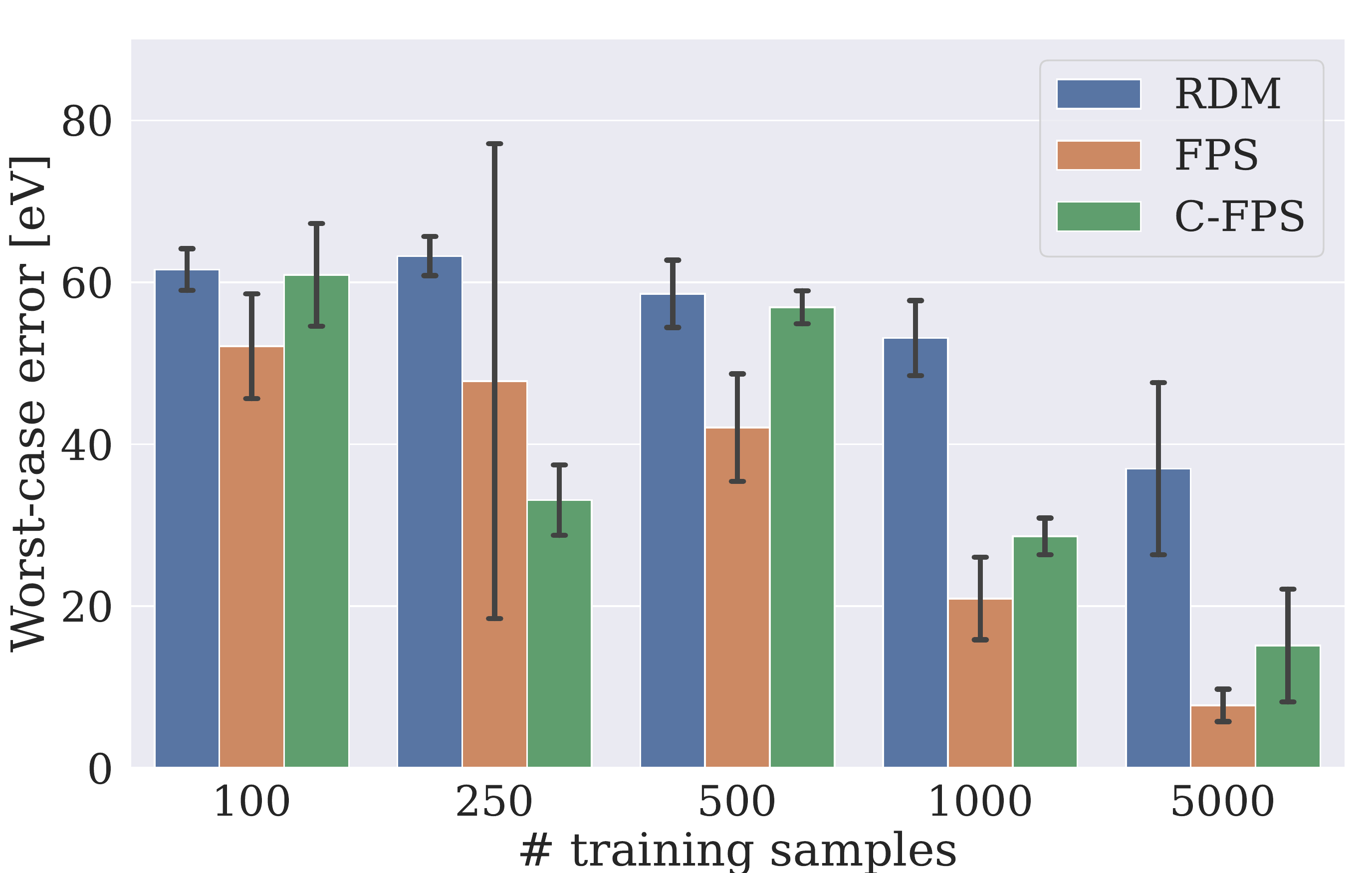}}
\caption{Results for spatial 3-hop convolution network}
\label{fig:Spatial_3_hop_convolution_network_results}
\end{figure}

In accordance with the previous sections, we train the method presented in Section \ref{subsec:ThreeHop} on subsets of QM9. The network consists of several updates by the spatial 3-hop convolution layer, followed by an aggregation layer to obtain graph features which are further processed by linear layers. During training we minimize the L1-loss function with respect to the models parameters with the Adam optimizer, use a learning rate of $2 \cdot 10^{-4}$ and a batch size of 32 molecules per iteration. We train each model for 500 epochs and choose the best model with respect to a validation set (20\% of the training set) for measuring the model performance on a test set. 

Apart from the model trained on 250 samples, we do not observe significant differences among the different sampling strategies when considering only the MAE (Fig. \ref{fig:Spatial_3_hop_convolution_network_results_a}). When training on 250 samples, the FPS-based methods appear to lead to an advantage and reduce the MAE in comparison to random sampling. For larger training sets random sampling seems to catch up and perform on par with FPS-based sampling. Considering the RMSE (Fig. \ref{fig:Spatial_3_hop_convolution_network_results_b}), our observations are somewhat different. In particular, we find FPS and C-FPS to outperform random sampling for most sizes of the training set. In line with the other methods, FPS reduces the worst-case error in comparison to random sampling (Fig. \ref{fig:Spatial_3_hop_convolution_network_results_c}). 

We observe comparatively large values for all metrics, especially for small training set sizes. For example, the MAE for 100 training samples obtained with FPS amounts to approximately \SI{5}{\eV}. This is around 4 times larger than what we measure for KRR and more than 10 times larger than the value of SchNet. This was to be expected, since both KRR and SchNet are relatively data efficient. We note that the good performace of SchNet is to be expected as it uses more features than the spatial 3-hop convolution network and especially uses the positions of the atoms (a powerful information which allows to compute the atomization energy explicitely). The KRR has the advantage of being a kernel method which has empirically shown to be effective in the realm of small datasets \cite{Pinheiro2021}. However, for larger training sets the relative difference between the methods becomes smaller: For 500 training samples, KRR yields only a two times smaller and SchNet only a 5 times smaller MAE. Moreover, the spatial 3-hop convolution network can benefit the most from larger datasets, i.e. we observe a significant improvement in performance whenever the size of the dataset is increased.

\subsection{Explanation} 
With GRDE framework from Section~\ref{subsec: GRDE} we now investigate the domain knowledge learned by the spatial 3-hop convolution network from Section \ref{subsec:ThreeHop} using the sampling strategies from Section \ref{sampling_strategies}.

\subsubsection*{Setup of the Experiments}
For the experiments, we utilize the spatial 3-hop convolution network from Section \ref{subsec:ThreeHop} that has been pre-trained using the sampling strategies from Section \ref{sampling_strategies}. More specifically, we compare explanations on the pre-trained model for the cases of random sampling (RDM) and diversity sampling (FPS) of 5000 samples as the training dataset. We fix the distortion measure $d$ as the $L^2$ distance for a regression task, and randomly initialize masks $S$. Furthermore, given the sparsity of the data, we also set a low value on $\lambda_A, \lambda_X=20$ (which corresponds to choosing 10-15\% of the non-zero elements in the respective masks) and set $(\varv_{S_A}, \varv_{S_X})$ to null. Since the QM9 dataset possesses edge features, we optimize $S_A=[S_{A_1}, S_{A_2},...,S_{A_h}]$ where $A_i$ is the adjacency matrix with respect to the edge feature $i$ $\forall i \in \{1,2, ..., h\}$; $h$ being the number of edge features. The results that follow are obtained as an average over 3 independent runs on 100 graphs randomly sampled from the respective test datasets. Since the setup produces positive relevance masks, i.e., the masks only obfuscate features that exist for each node/edge, and do not show the relevance of the lack of a feature for a node/edge), we aggregate and average the node- and edge-wise scores to obtain feature-wise scores. Furthermore, we offset the imbalance in the scores by weighting them with respect to the frequency of their occurrence over the sampled data. Our explanation query is as follows: \textit{For a randomly selected graph unseen by the pre-trained model, which features does the model consider important for its prediction?}

\begin{figure}[htb]
\centering
\subfloat[Node feature explanations]{\label{fig:X_results}\includegraphics[width=.48\linewidth]{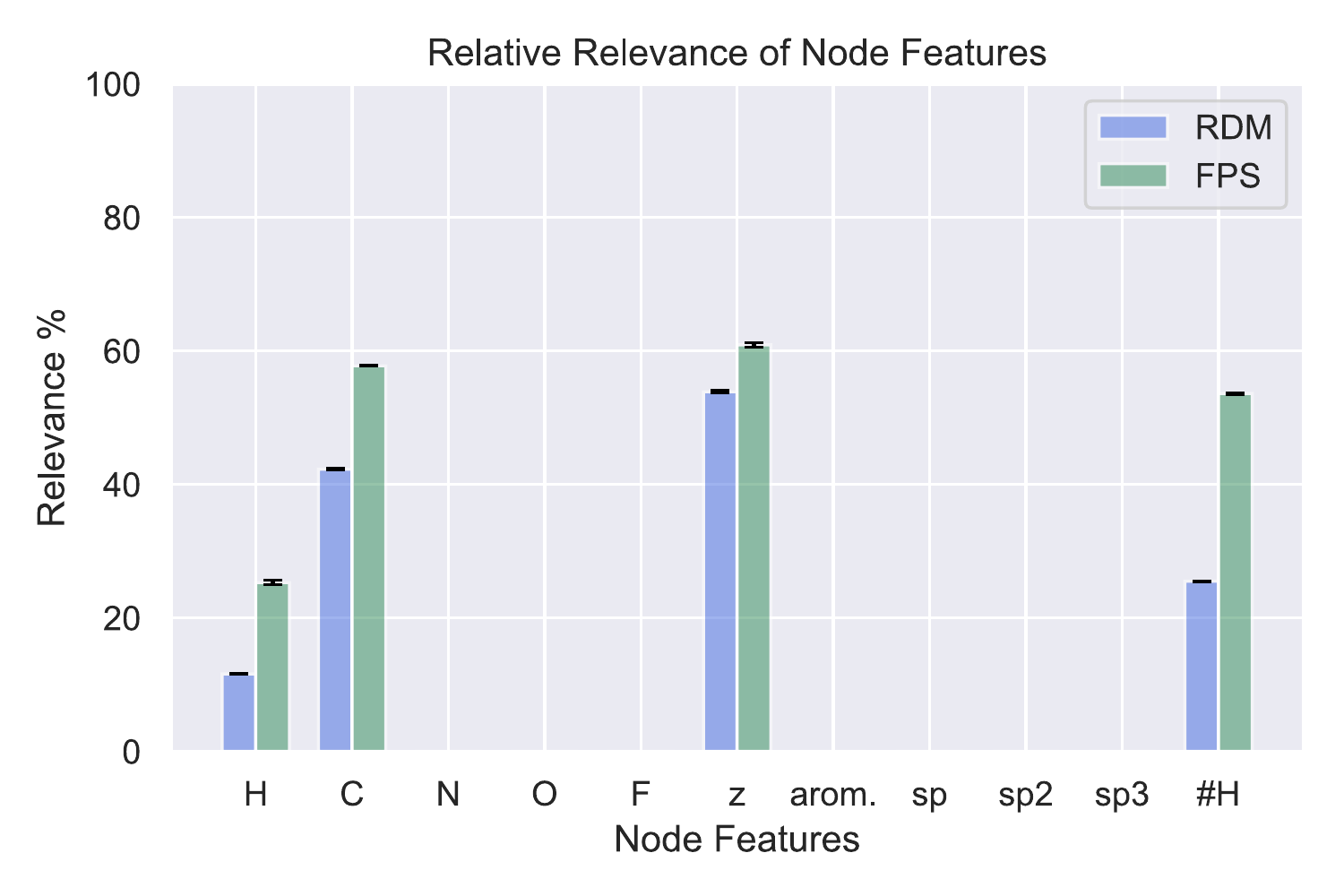}}\hfill
\subfloat[Edge feature explanations]{\label{fig:T_results}\includegraphics[width=.48\linewidth]{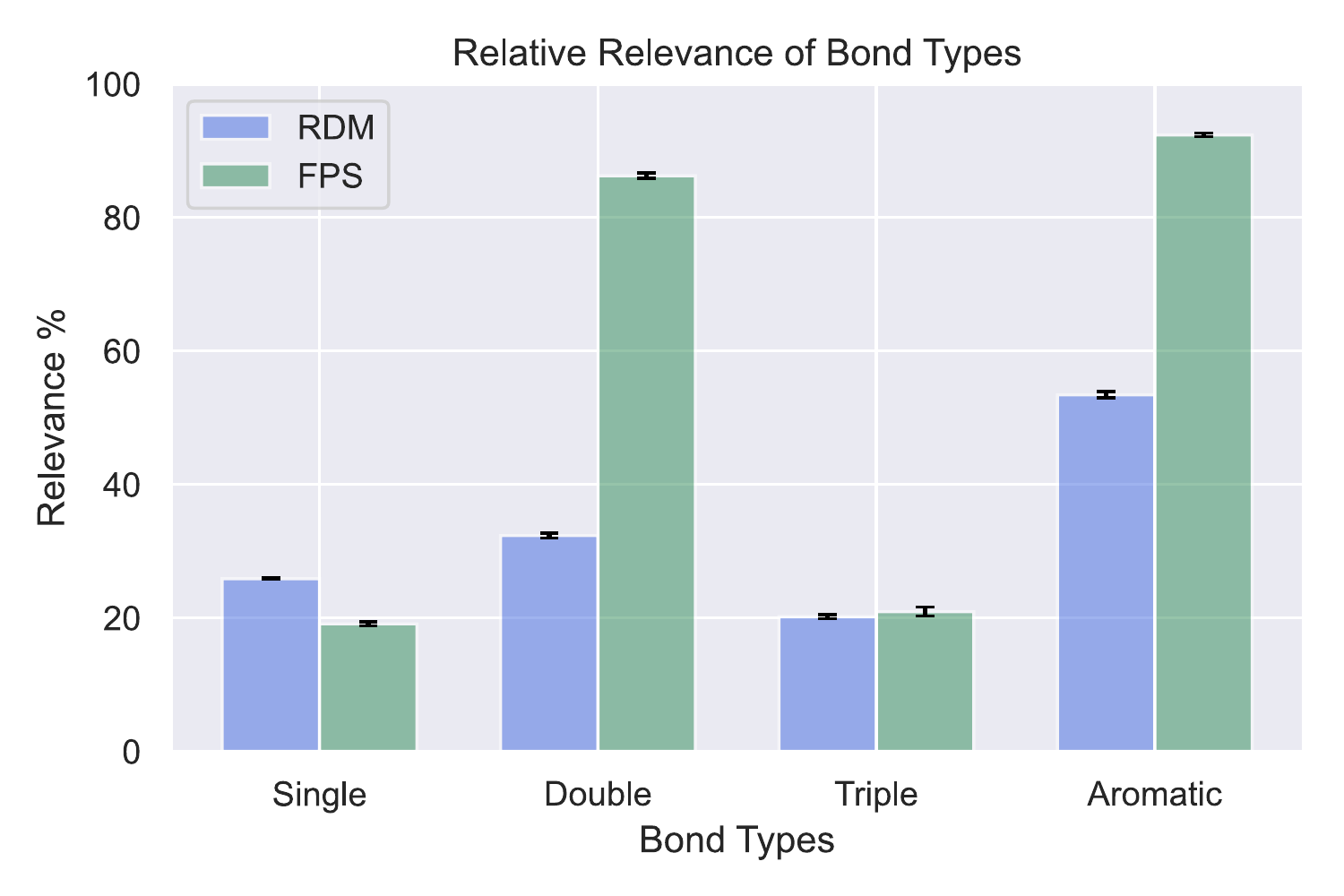}}
\caption{Results for feature-level GRDE.}
\label{fig:GRDE_results}
\end{figure}

\subsubsection*{Results}
From Fig. \ref{fig:X_results} and Fig. \ref{fig:T_results}, we see that the model trained using FPS places a stronger importance in both edge and node features than in the case of RDM, especially in the case of atomic numbers (Z), double bonds and aromatic rings. In comparison, single bonds, though the most frequent of the bond types in the sampled data are not considered as important. Furthermore, Fig. \ref{fig:X_results} shows that the model requires certain nodes to be specific atom types and have a certain type of neighborhood for its predictions, as can be seen from the atomic number (Z) as well as the importance of carbon (C) and the number of hydrogen atoms (\#H) surrounding the node in the case of FPS. In the case of RDM, we also have a clear indication that edge features are not as important as the node features, whereas this is significantly more balanced in the case of FPS. Finally, we find that, though there exist non-zero values for some node features in the sampled graphs such as in the case of Nitrogen (N) and Oxygen (O), GRDE does not attribute any importance to them. This implies that the model treats these features as noise and ignores them regardless of the sampling strategy used.


\section{Conclusion}

In this work, we employed three informed ML models to predict the atomization energy of molecules in the QM9 dataset. We used KRR  with a kernel obtained from molecular topological features, a geometry-based GNN (SchNet) and a topology-based GNN. We saw that maximizing molecular diversity in the training set selection process improves the accuracy and robustness of those methods. Our main finding is that by training topology-based ML models with sets of diverse molecules, we can significantly reduce their test maximum absolute error, thus increasing their robustness to distribution shifts. For SchNet, this effect was still observable but only for small training sets.
Moreover, by maximizing diversity in the training sets, we could substantially reduce the gap between the maximum absolute errors of a topology-based regression method as KRR and the SchNet, which is a geometry-based GNN. This work proposes only an empirical investigation, in the field of molecular property prediction, on the effects of maximizing molecular diversity in the training set selection. Ongoing research seeks to provide a theoretical foundation for the observed empirical results.

We believe that reducing the worst-case error is of great importance for applications that require a high degree of robustness but have limited budget for data generation. One example would be the application of Machine Learning Interatomic Potentials (MLIPs) for molecular dynamics simulations. In this scenario, the predictions of the model are used to integrate the equations of motion and to compute particle trajectories. Thus, large errors in the predictions could potentially lead to a failure of the simulation and techniques to prevent this are needed. Investigating this scenario in particular, could be a direction of future research.

\begin{acknowledgement}
This work was supported in part by the BMBF-project 05M20 MaGriDo (Mathematics for Machine Learning Methods for Graph-Based Data with Integrated Domain Knowledge) and in part by the Fraunhofer Cluster of
Excellence \guillemotright Cognitive Internet Technologies\guillemotleft.
\end{acknowledgement}
%

%
%
%
\bibliographystyle{plain}
\bibliography{references}

\begin{thebibliography}{10}

\bibitem{agarwal2023evaluating}
Chirag Agarwal, Owen Queen, Himabindu Lakkaraju, and Marinka Zitnik.
\newblock Evaluating explainability for graph neural networks.
\newblock {\em Scientific Data}, 10(1):144, 2023.

\bibitem{Agarwal2005}
P.~K. Agarwal, S.~Har-Peled, and K.~R. Varadarajan.
\newblock Geometric approximation via coresets.
\newblock In {\em Combinatorial and Computational Geometry}, volume~52, pages
  1--30. Cambridge University Press, 2005.

\bibitem{Barker2021}
James Barker, Laura-Sophie Berg, Jan Hamaekers, and Astrid Maass.
\newblock Rapid prescreening of organic compounds for redox flow batteries: A
  graph convolutional network for predicting reaction enthalpies from {SMILES}.
\newblock {\em Batteries {\&} Supercaps}, 4(9):1482--1490, jun 2021.

\bibitem{Batzner2022}
Simon Batzner, Albert Musaelian, Lixin Sun, Mario Geiger, Jonathan~P. Mailoa,
  Mordechai Kornbluth, Nicola Molinari, Tess~E. Smidt, and Boris Kozinsky.
\newblock E(3)-equivariant graph neural networks for data-efficient and
  accurate interatomic potentials.
\newblock {\em Nature Communications}, 13(1), may 2022.

\bibitem{Bochkarev2022}
Anton Bochkarev, Yury Lysogorskiy, Sarath Menon, Minaam Qamar, Matous Mrovec,
  and Ralf Drautz.
\newblock Efficient parametrization of the atomic cluster expansion.
\newblock {\em Phys. Rev. Materials}, 6:013804, Jan 2022.

\bibitem{brandstetter2022geometric}
Johannes Brandstetter, Rob Hesselink, Elise van~der Pol, Erik~J Bekkers, and
  Max Welling.
\newblock Geometric and physical quantities improve e(3) equivariant message
  passing.
\newblock In {\em International Conference on Learning Representations}, 2022.

\bibitem{Braverman2016NewFF}
Vladimir Braverman, Dan Feldman, and Harry Lang.
\newblock New frameworks for offline and streaming coreset constructions.
\newblock {\em ArXiv}, abs/1612.00889, 2016.

\bibitem{cai2021}
Shengze Cai, Zhicheng Wang, Sifan Wang, Paris Perdikaris, and George~Em
  Karniadakis.
\newblock Physics-informed neural networks for heat transfer problems.
\newblock {\em Journal of Heat Transfer}, 143(6), 2021.

\bibitem{cuomo2022}
Salvatore Cuomo, Vincenzo~Schiano Di~Cola, Fabio Giampaolo, Gianluigi Rozza,
  Maziar Raissi, and Francesco Piccialli.
\newblock Scientific machine learning through physics--informed neural
  networks: where we are and what’s next.
\newblock {\em Journal of Scientific Computing}, 92(3):88, 2022.

\bibitem{defferrard2016}
Micha\"{e}l Defferrard, Xavier Bresson, and Pierre Vandergheynst.
\newblock Convolutional neural networks on graphs with fast localized spectral
  filtering.
\newblock In {\em Proceedings of the 30th International Conference on Neural
  Information Processing Systems}, NIPS'16, page 3844–3852, Red Hook, NY,
  USA, 2016. Curran Associates Inc.

\bibitem{Gabor2021}
Volker~L. Deringer, Albert~P. Bartók, Noam Bernstein, David~M. Wilkins,
  Michele Ceriotti, and Gábor Csányi.
\newblock Gaussian process regression for materials and molecules.
\newblock {\em Chemical Reviews}, 121(16):10073--10141, 2021.
\newblock PMID: 34398616.

\bibitem{graphsvx}
Alexandre Duval and Fragkiskos~D Malliaros.
\newblock Graphsvx: Shapley value explanations for graph neural networks.
\newblock In {\em Joint European Conference on Machine Learning and Knowledge
  Discovery in Databases}, pages 302--318. Springer, 2021.

\bibitem{FPS}
Y.~Eldar, M.~Lindenbaum, M.~Porat, and Y.Y. Zeevi.
\newblock The farthest point strategy for progressive image sampling.
\newblock {\em {IEEE} Transactions on Image Processing}, 6(9):1305--1315, sep
  1997.

\bibitem{Feldman2019}
Dan Feldman.
\newblock Core-sets: Updated survey.
\newblock In {\em Sampling Techniques for Supervised or Unsupervised Tasks},
  pages 23--44. Springer International Publishing, oct 2019.

\bibitem{Fey2019}
Matthias Fey and Jan~E. Lenssen.
\newblock Fast graph representation learning with {PyTorch Geometric}.
\newblock In {\em ICLR Workshop on Representation Learning on Graphs and
  Manifolds}, 2019.

\bibitem{gasteiger_gemnet_2021}
Johannes Gasteiger, Chandan Yeshwanth, and Stephan G{\"u}nnemann.
\newblock Directional message passing on molecular graphs via synthetic
  coordinates.
\newblock In {\em Conference on Neural Information Processing Systems
  (NeurIPS)}, 2021.

\bibitem{gilmer2017}
Justin Gilmer, Samuel~S. Schoenholz, Patrick~F. Riley, Oriol Vinyals, and
  George~E. Dahl.
\newblock Neural message passing for quantum chemistry.
\newblock In {\em Proceedings of the 34th International Conference on Machine
  Learning - Volume 70}, ICML'17, page 1263–1272. JMLR.org, 2017.

\bibitem{gori2005}
M.~Gori, G.~Monfardini, and F.~Scarselli.
\newblock A new model for learning in graph domains.
\newblock In {\em Proceedings. 2005 IEEE International Joint Conference on
  Neural Networks}, volume~2, pages 729--734 vol. 2, 2005.

\bibitem{D2CP00268J}
Tim Gould and Stephen~G. Dale.
\newblock Poisoning density functional theory with benchmark sets of difficult
  systems.
\newblock {\em Phys. Chem. Chem. Phys.}, 24:6398--6403, 2022.

\bibitem{hernandez2022}
Quercus Hern{\'a}ndez, Alberto Bad{\'\i}as, Francisco Chinesta, and El{\'\i}as
  Cueto.
\newblock Thermodynamics-informed graph neural networks.
\newblock {\em arXiv preprint arXiv:2203.01874}, 2022.

\bibitem{graphlime}
Qiang Huang, Makoto Yamada, Yuan Tian, Dinesh Singh, and Yi~Chang.
\newblock Graphlime: Local interpretable model explanations for graph neural
  networks.
\newblock {\em IEEE Transactions on Knowledge and Data Engineering}, 2022.

\bibitem{gumbel_softmax}
Eric Jang, Shixiang Gu, and Ben Poole.
\newblock Categorical reparameterization with {G}umbel-softmax.
\newblock In {\em International Conference on Learning Representations}, 2017.

\bibitem{joerding1991}
Wayne~H Joerding and Jack~L Meador.
\newblock Encoding a priori information in feedforward networks.
\newblock {\em Neural Networks}, 4(6):847--856, 1991.

\bibitem{jorgensen2018}
{Peter Bj{\o}rn} J{\o}rgensen, {Karsten Wedel} Jacobsen, and {Mikkel
  N{\o}rgaard} Schmidt.
\newblock Neural message passing with edge updates for predicting properties of
  molecules and materials.
\newblock In {\em 32nd Conference on Neural Information Processing Systems,
  NIPS 2018}, 2018.

\bibitem{kim2022}
Yeonjoon Kim, Jaeyoung Cho, Nimal Naser, Sabari Kumar, Keunhong Jeong, Robert~L
  McCormick, Peter C~St John, and Seonah Kim.
\newblock Physics-informed graph neural networks for predicting cetane number
  with systematic data quality analysis.
\newblock {\em Proceedings of the Combustion Institute}, 2022.

\bibitem{kipf2017}
Thomas~N. Kipf and Max Welling.
\newblock Semi-supervised classification with graph convolutional networks.
\newblock In {\em 5th International Conference on Learning Representations,
  {ICLR} 2017}. OpenReview.net, 2017.
\newblock \url{https://openreview.net/forum?id=SJU4ayYgl}.

\bibitem{kolek}
Stefan Kolek, Duc~Anh Nguyen, Ron Levie, Joan Bruna, and Gitta Kutyniok.
\newblock A rate-distortion framework for explaining black-box model decisions.
\newblock {\em xxAI - Beyond Explainable AI}, page 91–115, 2022.

\bibitem{kolek2022explaining}
Stefan Kolek, Robert Windesheim, Hector~Andrade Loarca, Gitta Kutyniok, and Ron
  Levie.
\newblock Explaining image classifiers with multiscale directional image
  representation.
\newblock {\em arXiv preprint arXiv:2211.12857}, 2022.

\bibitem{kramer1992}
Mark~A Kramer, Michael~L Thompson, and Phiroz~M Bhagat.
\newblock Embedding theoretical models in neural networks.
\newblock In {\em 1992 American Control Conference}, pages 475--479. IEEE,
  1992.

\bibitem{krause2014submodular}
Andreas Krause and Daniel Golovin.
\newblock Submodular function maximization.
\newblock {\em Tractability}, 3:71--104, 2014.

\bibitem{Kulik_2022}
H~J Kulik, T~Hammerschmidt, J~Schmidt, S~Botti, M~A~L Marques, M~Boley,
  M~Scheffler, M~Todorovi{\'{c}}, P~Rinke, C~Oses, A~Smolyanyuk, S~Curtarolo,
  A~Tkatchenko, A~P Bart{\'{o}}k, S~Manzhos, M~Ihara, T~Carrington, J~Behler,
  O~Isayev, M~Veit, A~Grisafi, J~Nigam, M~Ceriotti, K~T Schütt, J~Westermayr,
  M~Gastegger, R~J Maurer, B~Kalita, K~Burke, R~Nagai, R~Akashi, O~Sugino,
  J~Hermann, F~No{\'{e}}, S~Pilati, C~Draxl, M~Kuban, S~Rigamonti, M~Scheidgen,
  M~Esters, D~Hicks, C~Toher, P~V Balachandran, I~Tamblyn, S~Whitelam,
  C~Bellinger, and L~M Ghiringhelli.
\newblock Roadmap on machine learning in electronic structure.
\newblock {\em Electronic Structure}, 4(2):023004, jun 2022.

\bibitem{Kung2014}
S.~Y. Kung.
\newblock {\em Kernel Methods and Machine Learning}.
\newblock Cambridge University Press, apr 2014.

\bibitem{RDKIT}
G.~Landrum.
\newblock Rdkit:.
\newblock {\em Open-source cheminformatics}, 2012.

\bibitem{li2015}
Yujia Li, Daniel Tarlow, Marc Brockschmidt, and Richard Zemel.
\newblock Gated graph sequence neural networks.
\newblock In Yoshua Bengio and Yann LeCun, editors, {\em 4th International
  Conference on Learning Representations, {ICLR} 2016, San Juan, Puerto Rico,
  May 2-4, 2016, Conference Track Proceedings}, 2016.

\bibitem{liu2020}
Zhiyuan Liu and Jie Zhou.
\newblock Introduction to graph neural networks.
\newblock {\em Synthesis Lectures on Artificial Intelligence and Machine
  Learning}, 14(2):1--127, 2020.

\bibitem{pgex}
Dongsheng Luo, Wei Cheng, Dongkuan Xu, Wenchao Yu, Bo~Zong, Haifeng Chen, and
  Xiang Zhang.
\newblock Parameterized explainer for graph neural network.
\newblock {\em Advances in neural information processing systems},
  33:19620--19631, 2020.

\bibitem{Ma2015}
Junshui Ma, Robert~P. Sheridan, Andy Liaw, George~E. Dahl, and Vladimir
  Svetnik.
\newblock Deep neural nets as a method for quantitative
  structure{\textendash}activity relationships.
\newblock {\em Journal of Chemical Information and Modeling}, 55(2):263--274,
  feb 2015.

\bibitem{macdonald2019}
Jan MacDonald, Stephan W{\"a}ldchen, Sascha Hauch, and Gitta Kutyniok.
\newblock A rate-distortion framework for explaining neural network decisions.
\newblock {\em arXiv preprint arXiv:1905.11092}, 2019.

\bibitem{concrete}
Chris~J Maddison, Andriy Mnih, and Yee~Whye Teh.
\newblock The concrete distribution: A continuous relaxation of discrete random
  variables.
\newblock In {\em International Conference on Learning Representations}, 2017.

\bibitem{gcexplainer}
Lucie~Charlotte Magister, Dmitry Kazhdan, Vikash Singh, and Pietro Li{\`o}.
\newblock {GCExplainer}: Human-in-the-loop concept-based explanations for graph
  neural networks.
\newblock In {\em 3rd ICML Workshop on Human in the Loop Learning}, 2021.
\newblock arXiv preprint arXiv:2107.11889.

\bibitem{Mahoney2009}
Michael~W. Mahoney and Petros Drineas.
\newblock {CUR} matrix decompositions for improved data analysis.
\newblock {\em Proceedings of the National Academy of Sciences},
  106(3):697--702, jan 2009.

\bibitem{Mordred}
Hirotomo Moriwaki, Yu-Shi Tian, Norihito Kawashita, and Tatsuya Takagi.
\newblock Mordred: A molecular descriptor calculator.
\newblock {\em Journal of Cheminformatics}, 10(1), feb 2018.

\bibitem{Mueller2016}
Tim Mueller, Aaron~Gilad Kusne, and Rampi Ramprasad.
\newblock Machine learning in materials science.
\newblock In {\em Reviews in Computational Chemistry}, pages 186--273. John
  Wiley {\&} Sons, Inc, may 2016.

\bibitem{scikit}
Fabian Pedregosa, Ga{\"e}l Varoquaux, Alexandre Gramfort, Vincent Michel,
  Bertrand Thirion, Olivier Grisel, Mathieu Blondel, Peter Prettenhofer, Ron
  Weiss, Vincent Dubourg, et~al.
\newblock Scikit-learn: Machine learning in python.
\newblock {\em Journal of Machine Learning Research}, 12(Oct):2825--2830, 2011.

\bibitem{Pinheiro2021}
Max Pinheiro, Fuchun Ge, Nicolas Ferré, Pavlo~O. Dral, and Mario Barbatti.
\newblock Choosing the right molecular machine learning potential.
\newblock {\em Chem. Sci.}, 12:14396--14413, 2021.

\bibitem{pope}
Phillip~E. Pope, Soheil Kolouri, Mohammad Rostami, Charles~E. Martin, and Heiko
  Hoffmann.
\newblock Explainability methods for graph convolutional neural networks.
\newblock In {\em 2019 IEEE/CVF Conference on Computer Vision and Pattern
  Recognition (CVPR)}, pages 10764--10773, 2019.

\bibitem{ramakrishnan2014quantum}
Raghunathan Ramakrishnan, Pavlo~O Dral, Matthias Rupp, and O~Anatole von
  Lilienfeld.
\newblock Quantum chemistry structures and properties of 134 kilo molecules.
\newblock {\em Scientific Data}, 1, 2014.

\bibitem{Roscher.Bohn.Duarte.ea:2019}
Ribana Roscher, Bastian Bohn, Marco~F. Duarte, and Jochen Garcke.
\newblock {Explainable Machine Learning for Scientific Insights and
  Discoveries}.
\newblock {\em IEEE Access}, 8(1):42200--42216, 2020.

\bibitem{Ruddigkeit2012}
Lars Ruddigkeit, Ruud van Deursen, Lorenz~C. Blum, and Jean-Louis Reymond.
\newblock Enumeration of 166 billion organic small molecules in the chemical
  universe database gdb-17.
\newblock {\em Journal of Chemical Information and Modeling},
  52(11):2864--2875, 2012.
\newblock PMID: 23088335.

\bibitem{scarselli2009}
Franco Scarselli, Marco Gori, Ah~Chung Tsoi, Markus Hagenbuchner, and Gabriele
  Monfardini.
\newblock The graph neural network model.
\newblock {\em IEEE Transactions on Neural Networks}, 20(1):61--80, 2009.

\bibitem{graphmask}
Michael~Sejr Schlichtkrull, Nicola~De Cao, and Ivan Titov.
\newblock Interpreting graph neural networks for {\{}nlp{\}} with
  differentiable edge masking.
\newblock In {\em International Conference on Learning Representations}, 2021.
\newblock \url{https://openreview.net/forum?id=WznmQa42ZAx}.

\bibitem{Schuett2017}
K.~T. Sch\"{u}tt, P.-J. Kindermans, H.~E. Sauceda, S.~Chmiela, A.~Tkatchenko,
  and K.-R. M\"{u}ller.
\newblock Schnet: A continuous-filter convolutional neural network for modeling
  quantum interactions.
\newblock In {\em Proceedings of the 31st International Conference on Neural
  Information Processing Systems}, NIPS'17, page 992–1002, Red Hook, NY, USA,
  2017. Curran Associates Inc.

\bibitem{shukla2020}
Khemraj Shukla, Patricio~Clark Di~Leoni, James Blackshire, Daniel Sparkman, and
  George~Em Karniadakis.
\newblock Physics-informed neural network for ultrasound nondestructive
  quantification of surface breaking cracks.
\newblock {\em Journal of Nondestructive Evaluation}, 39:1--20, 2020.

\bibitem{smilkov2017smoothgrad}
Daniel Smilkov, Nikhil Thorat, Been Kim, Fernanda Vi{\'e}gas, and Martin
  Wattenberg.
\newblock Smoothgrad: removing noise by adding noise.
\newblock {\em arXiv preprint arXiv:1706.03825}, 2017.

\bibitem{Sutton2020}
Christopher Sutton, Mario Boley, Luca~M. Ghiringhelli, Matthias Rupp, Jilles
  Vreeken, and Matthias Scheffler.
\newblock {Identifying domains of applicability of machine learning models for
  materials science}.
\newblock {\em Nature Communications}, 11(1):4428, sep 2020.

\bibitem{todeschini2009molecular}
R~Todeschini and V~Consonni.
\newblock {\em Molecular Descriptors for Chemoinformatics}.
\newblock Wiley-VCH, 2009.

\bibitem{velivckovic2017}
Petar Veli{\v{c}}kovi{\'c}, Guillem Cucurull, Arantxa Casanova, Adriana Romero,
  Pietro Lio, and Yoshua Bengio.
\newblock Graph attention networks.
\newblock In {\em International Conference on Learning Representations}, 2018.
\newblock https://openreview.net/forum?id=rJXMpikCZ.

\bibitem{TaxonomyML2021}
Laura von Rueden, Sebastian Mayer, Katharina Beckh, Bogdan Georgiev, Sven
  Giesselbach, Raoul Heese, Birgit Kirsch, Julius Pfrommer, Annika Pick,
  Rajkumar Ramamurthy, Michał Walczak, Jochen Garcke, Christian Bauckhage, and
  Jannis Schuecker.
\newblock {Informed Machine Learning - A Taxonomy and Survey of Integrating
  Knowledge into Learning Systems}.
\newblock {\em IEEE Transactions on Knowledge and Data Engineering},
  35(1):614--633, 2023.

\bibitem{SMILES}
David Weininger.
\newblock {SMILES}, a chemical language and information system. 1. introduction
  to methodology and encoding rules.
\newblock {\em Journal of Chemical Information and Modeling}, 28(1):31--36, feb
  1988.

\bibitem{gnnexplainer}
Zhitao Ying, Dylan Bourgeois, Jiaxuan You, Marinka Zitnik, and Jure Leskovec.
\newblock Gnnexplainer: Generating explanations for graph neural networks.
\newblock {\em Advances in neural information processing systems}, 32, 2019.

\bibitem{Yu06}
Kai Yu, Jinbo Bi, and Volker Tresp.
\newblock Active learning via transductive experimental design.
\newblock In {\em Proceedings of the 23rd International Conference on Machine
  Learning}, ICML '06, page 1081–1088, New York, NY, USA, 2006. Association
  for Computing Machinery.

\bibitem{xgnn}
Hao Yuan, Jiliang Tang, Xia Hu, and Shuiwang Ji.
\newblock Xgnn: Towards model-level explanations of graph neural networks.
\newblock In {\em Proceedings of the 26th ACM SIGKDD International Conference
  on Knowledge Discovery \& Data Mining}, pages 430--438, 2020.

\bibitem{yuan2022explainability}
Hao Yuan, Haiyang Yu, Shurui Gui, and Shuiwang Ji.
\newblock Explainability in graph neural networks: A taxonomic survey.
\newblock {\em IEEE Transactions on Pattern Analysis and Machine Intelligence},
  2022.

\bibitem{subgraphx}
Hao Yuan, Haiyang Yu, Jie Wang, Kang Li, and Shuiwang Ji.
\newblock On explainability of graph neural networks via subgraph explorations.
\newblock In {\em International Conference on Machine Learning}, pages
  12241--12252. PMLR, 2021.

\bibitem{Zaverkin2022}
Viktor Zaverkin, David Holzmüller, Ingo Steinwart, and Johannes Kästner.
\newblock Exploring chemical and conformational spaces by batch mode deep
  active learning.
\newblock {\em Digital Discovery}, 2022.

\end{thebibliography}

\end{document}